\documentclass[review,hidelinks,onefignum,onetabnum]{siamart251216}

\usepackage{lipsum}
\usepackage{amsfonts}
\usepackage{graphicx}
\usepackage{epstopdf}
\usepackage{algorithmic}
\usepackage{bm}
\usepackage{cases}
\usepackage{subfigure}

\newtheorem{Thm}{Theorem}[section]

\newtheorem{Pro}{Proposition}[section]
\newtheorem{Remark}{Remark}[section]
\renewcommand{\figurename}{Fig.}

\ifpdf
\DeclareGraphicsExtensions{.eps,.pdf,.png,.jpg}
\else
\DeclareGraphicsExtensions{.eps}
\fi

\newsiamremark{remark}{Remark}
\newsiamremark{hypothesis}{Hypothesis}
\crefname{hypothesis}{Hypothesis}{Hypotheses}
\newsiamremark{fact}{Fact}
\crefname{fact}{Fact}{Facts}

\headers{Microscopic to macroscopic pedestrian models}{L. Yang, H. Yu, and J. Du}

\title{From microscopic social force models to macroscopic continuum models for pedestrian flow \thanks{Submitted to the editors DATE. 
	   \funding{The work of L. Yang is supported under Grants NSFC 12271499. The work of H. Yu is supported under Grants NSFC 12271288 and 12571469, and in part by the 111 Project (No. D23017), and Program for Science and Technology Innovative Research Team in Higher Educational Institutions of Hunan Province of China. The work of J. Du is supported by the National Natural Science Foundation of China (Grant No. 12571425), Natural Science Foundation of Shanghai, China (Grant No. 24ZR1417600), and Shanghai Young Academic Program of Eastern Talent Plan. }}}

\author{Liangze Yang\thanks{School of Mathematical Sciences, Anhui University, Hefei, Anhui, 230601, China
		(\email{lzeyang@ahu.edu.cn}).}
	\and Hui Yu\thanks{School of Mathematics and Computational Science, Xiangtan University, Xiangtan, 411105, China 
		(\email{huiyu@xtu.edu.cn}).}
	\and Jie Du\thanks{School of Mathematical Sciences, Key Laboratory of MEA (Ministry of Education) $\&$ Shanghai Key Laboratory of PMMP, East China Normal University, Shanghai, 200241, China
		(\email{jdu@math.ecnu.edu.cn}, The corresponding author).}
}
\usepackage{amsopn}

\ifpdf
\hypersetup{
  pdftitle={Microscopic to macroscopic pedestrian models},
  pdfauthor={L. Yang, H. Yu, and J. Du}
}
\fi

\begin{document}
\nolinenumbers
\maketitle

\begin{abstract}
The pedestrian flow is one of the most complex systems, involving large populations of interacting agents. Models at microscopic and macroscopic scales offer different advantages for studying related problems. In general, microscopic models can describe interaction forces at the individual level. Macroscopic models, on the other hand, provide analytical insights into global interactions and long-term overall dynamics, along with efficient numerical simulations and predictions. However, the relationship between models at different scales has rarely been explored. In this study, based on the original microscopic social force model with a reactive optimal route choice strategy, we first derive kinetic equations at the mesoscopic level. 
By varying the interaction force in different scenarios, we then derive several continuum models at the macroscopic level. Finally, numerical examples are given to evaluate the behaviors of the social force model and our continuum models.   
\end{abstract}



\section{Introduction}
Due to the importance of safety management, pedestrian and crowd dynamics have received considerable attention, especially for large-scale and high-density crowds. Over the past two decades, crowd disasters have caused thousands of deaths worldwide, such as the 2010 Love Parade disaster (652 injuries), 2015 Saudi Arabia Hajj disaster (2431 deaths) and the 2022 Seoul Halloween crush (156 deaths). 
Various crowd forces significantly influence pedestrian dynamics, which can be attributed to two main categories: local interacting forces and global forces \cite{liang2021continuum}. Experiment results show that the maximum measured interacting force generated by an individual ranges from 162 $N$  for a fit young female adult to 600 $N$ for a super-fit young male adult \cite{dickie1993spectator}. In contrast, the global force refers to the aggregated force that emerges in a dense crowd. Under high density conditions, frequent physical contact between pedestrians allows forces to be transferred and accumulated throughout the crowd.
This global force can escalate rapidly, further amplified by surrounding obstacles and the friction between people and the ground. The forensic study shows that the magnitude of such global force can reach 6000 $N$, and the work \cite{helbing2012crowd} by Helbing and Mukerji shows that the global force is a critical characteristic of panicked crowds.

Dynamic pedestrian flow models serve as a valuable tool for comprehending, replicating, and predicting the behavior of pedestrian flow. Existing models can generally be classified into three descriptive scales  \cite{albi2019vehicular,bellomo2012modeling,duives2013state}: microscopic, mesoscopic, and macroscopic models. Classic microscopic models are the most widely used models, such as individual-based models or cellular automaton-based models. The movements of pedestrians are described by the physical and psychological interactions between individuals. Such models are governed by Newton-type equations \cite{bellomo2011modeling,helbing2000simulating,kaji2017cellular}. At the mesoscopic scale, the positions and velocities of pedestrians are characterized using probability distributions, and the models are governed by mean field equations \cite{bellomo2015multiscale}. At the macroscopic scale, pedestrian dynamics are described as continuum flow, and variables such as density and velocity can be represented by mathematical functions. As a result, the models are usually governed by fluid dynamic equations, such as conservation laws that describe the conservation of mass, momentum, and/or energy of pedestrian flow \cite{etikyala2014particle,hughes2002continuum}.

To quantitatively study local interacting forces, microscopic models determine directly various interacting forces from a microscopic perspective, and are able to simulate a great deal of rich phenomena, such as the lane formation. The social force model (SFM)  \cite{helbing1995social} is one of the most classic microscopic models that consider physical contact forces and psychological tendency forces between individuals, aiming to characterize their attraction, repulsion and panic.
In recent two decades, the SFM has been widely used to understand pedestrian dynamics at the microscopic level, and numerous extended models have been established for applications in different fields, including human behavior modeling, traffic flow analysis, and architectural design \cite{lakoba2005modifications,yang2014guided}. 
However, these models are more suitable for simulating pedestrian movements at low-density levels, and may consume a huge amount of computational costs for a high density pedestrian system. Moreover, the existing microscopic models are difficult to accurately describe the global force at the microscopic level. 

For large-scale and high-density crowds, each individual follows similar principles of dynamics.
When there is no need to distinguish each individual, and the focuses are on the overall crowd characteristics, macroscopic models show their advantages, such as the well-known Lighthill–Whitham–Richards (LWR) model \cite{lighthill1955kinematic,richards1956shock} and Payne–Whitham (PW) model \cite{whitham1974linear}. 
Recently, macroscopic models have been further refined to reproduce the dense, high-pressure
crowds emerged during crowd disasters, explicitly considering the effect of panic on pedestrian
behaviours in which the global force is considered.
And the stop-and-go wave and turbulence have been observed in numerical results \cite{liang2021continuum,liang2024modelling}. 
 However, because of the difficulty of parameter calibration, the pressure setting in these models is often questioned by experimentalists. 

Microscopic models, which explicitly track the trajectories and interactions of individual pedestrians, are generally more suitable for low-density scenarios where individual behaviors and discrete interactions dominate the dynamics. 
In contrast, macroscopic models describe pedestrian flows in terms of aggregate variables such as density and average velocity, making them more efficient and accurate in high-density situations where collective motion and continuum effects prevail. 
However, most macroscopic models are formulated directly from a global or phenomelogical perspective without being rigorously derived from underlying individual-based dynamics. 
When the density gradually increases from a low level to a high level, the mathematical connections between existing models at different scales remain insufficiently understood. 
This lack of a unified framework hampers our ability to model transitional regimes consistently. 
In related research fields, bottom-up modeling strategies—starting from microscopic or mesoscopic descriptions and systematically deriving macroscopic equations—have been developed to bridge such scale gaps, for instance in kinetic theory of gases \cite{cercignani2013mathematical}, traffic flow modeling \cite{piccoli2013vehicular}, and active matter systems\cite{marchetti2013hydrodynamics}. These approaches offer a promising direction for developing macroscopic pedestrian models that accurately represent microscopic mechanisms across a wide range of densities. 

In this paper, we would like to employ the bottom-up modeling strategy. Starting from microscopic models, we then derive the kinetic equations at the mesoscopic level as the density increases, and finally get the macroscopic models for high-density conditions. Thus, we can establish the ground connections between models at different scales. 
This modeling strategy provides several advantages. On the one hand, the microscopic model can be used as a ground truth to calibrate parameters in the macroscopic model, and provide some insights of the physical meaning of these parameters. 
On the other hand, the macroscopic model can give a theoretical study and efficient numerical simulations to evaluate and adjust the global force for the microscopic involution. 



We will start from the original SFM \cite{helbing1995social}, and assume that the pedestrian's desired velocity is governed by the reactive dynamic user-optimal principle \cite{huang2009revisiting,yang2019modeling}.
Thus, the desired walking direction satisfies an eikonal equation \cite{huang2009revisiting,jiang2020dynamic}. 
Using this novel coupling system, we derive mesoscopic kinetic equations and macroscopic continuum equations by using the kinetic theory under different assumptions of the attraction and repulsion force, including the dimensionalized case and the hydrodynamic scaling case. 
We will also provide numerical examples to compare the microscopic social force model and the new derived continuum model, and thus show the evaluation of our new model. The third-order Runge-Kutta method is applied to solve the social force model. The efficient fifth-order weighted essentially non-oscillatory (WENO) scheme \cite{cockburn1998essentially,jiang1996efficient,liu1994weighted} is used for solving the continuum equation. We also adopt the fast sweeping method \cite{xiong2011high} for solving the eikonal equation and the third-order total variation diminishing (TVD) Runge-Kutta for time discretizations. 

The remaining part of the paper is organized as follows. 
Section \ref{secMicroModels} introduces the microscopic social force model coupled with an optimal route choice strategy, and analyzes the time-asymptotic flocking behavior. 
Sections \ref{secMeanFieldModels} and \ref{secMacroModels} use the kinetic theory to derive mean-field models at the mesoscopic level, and then provide the final macroscopic model, respectively. 
Section \ref{secNumericalExamples} presents the designed numerical examples to compare results between the microscopic model and the macroscopic model, and provides discussions of the macroscopic models. Section \ref{secConclusion} shows concluding remarks and future works.

\section{Microscopic social force model}\label{secMicroModels}
The microscopic social force model is developed to describe the pedestrian flow. In this model, pedestrians are treated as self-driven particles, and pedestrians' behavior is assumed to be influenced by a mixture of socio-psychological and physical forces. In this section, we briefly introduce the general social force model which is integrated with an optimal route choice strategy \cite{huang2009revisiting} for movement towards specific destinations or exits. We will also analyze the time-asymptotic flocking behavior.

\subsection{Social force model coupled with an optimal route choice} 

For a two-dimensional interacting particle system comprising $N$ individuals within a confined region $\Omega \in \mathbb{R}^2$, the motion of the $i$-th ($i=1, 2, \ldots, N$) pedestrian is governed by 
\begin{subequations}
\begin{numcases}{}
    \hspace{0.45cm}\frac{d\bm{x}_{i}(t)}{dt}=\bm{v}_{i}(t), \\
    m_{i}\frac{d\bm{v}_{i}(t)}{dt}=m_{i}\frac{\bm{u}_{i}^{\rm e}(t)-\bm{v}_{i}(t)}{\tau}+\sum_{j\ne i}\bm{f}_{ij}+\sum_{\rm walls}\bm{f}_{i}^{\rm wall},
\end{numcases} 
\label{sf}
\end{subequations}
where $\bm{x}_{i}(t)$ and $\bm{v}_{i}(t)$ are the position and the real velocity of the $i$-th  pedestrian at time $t$, and $m_{i}$ is the mass of pedestrian $i$.
Here $\tau>0$ is the constant relaxation time. 

We use $\bm{u}_{i}^{\rm e}(t)$ to denote the desired or expected velocity vector of pedestrian $i$ at time $t$, which depends on the pedestrian density. We assume that all pedestrians follow the same strategy to determine their desired velocity. In other words, we have 
\[
\bm{u}_{i}^{\rm e}(t)=\bm{u}^{\rm e}(\bm{x}_{i}(t),t) = U^{\rm e}\big(\bm{x}_{i}(t),t\big)\bm{e}\big(\bm{x}_{i}(t),t\big),
\]
where $U^{\rm e}(\bm{x},t)$ is the desired speed magnitude of pedestrians at location $\bm{x}$ and time $t$, and $\bm{e}(\bm{x},t)$ is the desired moving direction to be specified later according to the optimal route choice strategy. Here the desired speed is computed by 
\begin{align}\label{eq_desired_speed}
   U^{\rm e}(\bm{x},t)=U_{f}\exp{\Big(-\beta\rho_{\rm loc}(\bm{x},t)\Big)},
\end{align}
where $U_{f}>0$ is the constant free-flow speed, and $\beta$ is a model parameter for scaling the local density function
$\rho_{\rm loc}(\bm{x},t)$. 
The nonnegative density function $\rho_{\rm loc}(\bm{x},t)$ is to evaluate the crowdedness level at location $\bm{x}$ and time $t$, measured by 
\begin{align}
    \rho_{\rm loc}(\bm{x},t)=\sum_{1\leq j\leq N}w\Big(\|\bm{x}_{j}(t)-\bm{x}\|\Big),
    \label{eq3}
\end{align}
where $w(r)$ is a Gaussian distance-dependent weight function given by
\begin{align}
    w(r)=\frac{1}{\pi R^2}\exp{\left(-\frac{r^2}{R^2}\right)}.
    \label{eq4}
\end{align}
Here $R>0$ is a measurement parameter for the interaction range among pedestrians. 
For the path choice strategy, we assume pedestrians always choose routes to minimize their instantaneous travel cost and change moving directions in a reactive manner \cite{hughes2002continuum,huang2009revisiting,jiang2020dynamic}.  
The desired moving direction satisfies the following equation:
\begin{align}\label{eq_desired_direction}
    \bm{e}(\bm{x},t) =-\frac{\nabla \varphi(\bm{x},t)}{\|\nabla \varphi(\bm{x},t)\|},
\end{align}
where $\varphi(\bm{x},t)$ denotes the travel cost potential for a pedestrian moving from position $\bm{x}$ at time $t$ to the given destination or exit,  governed by the following eikonal equation:
\begin{subequations}
\begin{numcases}{}
    \hspace{0.45cm}\|\nabla \varphi(\bm{x},t)\|=c(\bm{x},t), \\
   \varphi(\bm{x},t)=\varphi_c(t), \ \ \bm{x}\in \Gamma_{c}.
\end{numcases} 
\end{subequations}
Here $c(\bm{x},t)$ is the given local travel cost per unit distance at location $\bm{x}$ and time $t$, and $\varphi_c(t)$ is the travel cost of leaving the exit $\Gamma_{c}$. For simplicity of notations, we abbreviate $U^{\rm e}(\bm{x},t)$ and $\bm{e}(\bm{x},t)$ as $U^{\rm e}(\bm{x})$ and $\bm{e}(\bm{x})$ in the following part of this paper. 

\begin{figure}[htbp]
	\centering
	\includegraphics[height=0.36\textwidth]{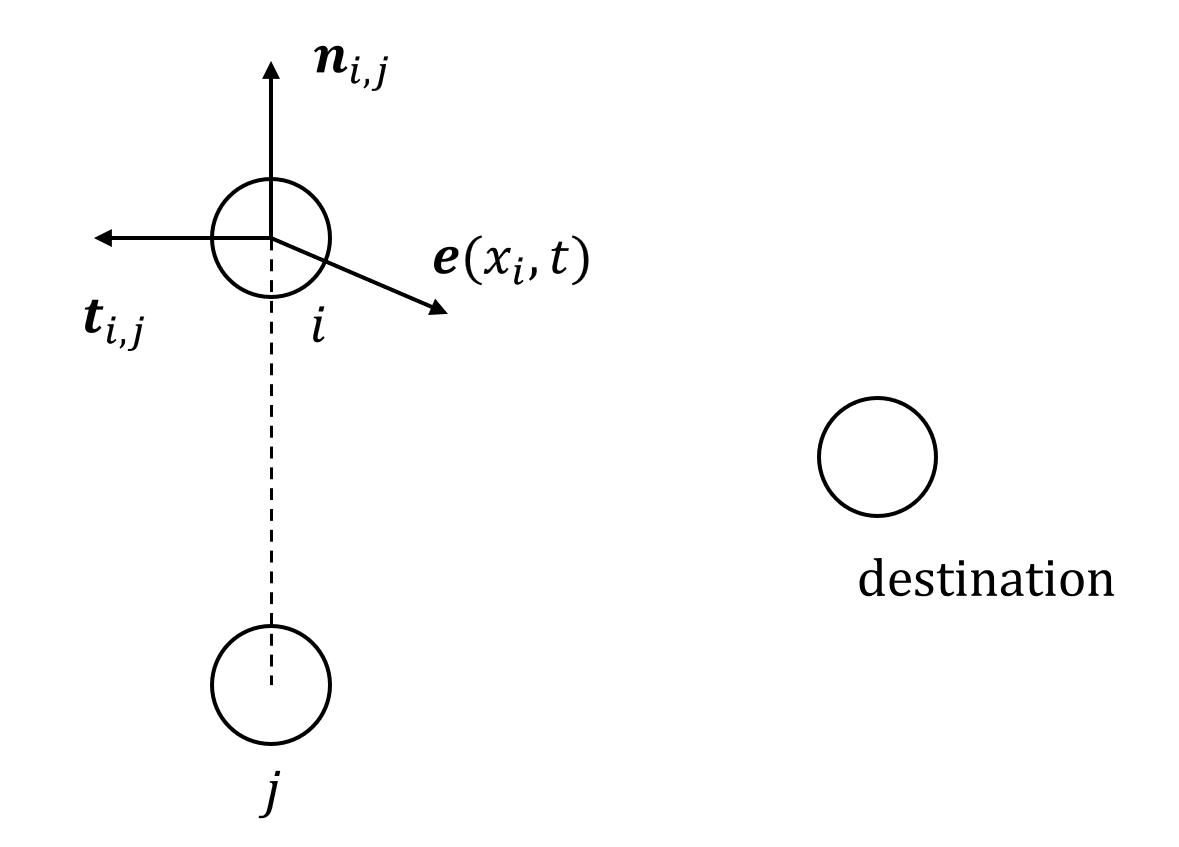}
	\caption{Diagram of the direction in the social force model.}
	\label{fig_force}
\end{figure}

In (\ref{sf}b), $\bm{f}_{i,j}$ is the interaction repulsion force between pedestrian $i$ and $j$: 
\begin{align*}
    \bm{f}_{i,j}=\left[A_{i}\exp\left(\frac{r_{i,j}-d_{i,j}}{B_{i}}\right)+kg(r_{i,j}-d_{i,j})\right]\bm{n}_{i,j}+\kappa g(r_{i,j}-d_{i,j})\langle\bm{v}_{j}-\bm{v}_{i}, \bm{t}_{i,j}\rangle \bm{t}_{i,j},
\end{align*}
where $A_{i}$ and $B_{i}$ are positive constants. Parameters $k$ and $\kappa$ are constants, usually large enough to ensure the volume exclusion of pedestrians. 
Let $r_{i}$ be the radius of pedestrian $i$. Then 
$$r_{i,j}=r_{i}+r_{j},\qquad d_{i,j}= \Vert \bm{x}_{i}-\bm{x}_{j}\Vert$$
are the minimum distance for pedestrians $i$ and $j$ to be comfortable and the real distance between them. 
The unit vector pointing from pedestrian $j$ to $i$ is given by 
\[
\bm{n}_{i,j}=\left(n_{i,j}^{1}, n_{i,j}^{2}\right)=\frac{\bm x_{i}-\bm x_{j}}{d_{i,j}}.
\]
Denote the tangential direction as $\bm{t}_{i,j}=\left(-n_{i,j}^{2},n_{i,j}^{1}\right)$.
The diagram of $\bm{n}_{i,j}$ and $\bm{t}_{i,j}$ is shown in Figure \ref{fig_force}. 
We use $\langle \cdot, \cdot \rangle$ to denote the dot product of two vectors.
Then $\langle\bm{v}_{j}-\bm{v}_{i}, \bm{t}_{i,j}\rangle$ becomes the velocity difference along the tangential direction. We adopt the nonnegative function $g(\gamma)$ to model the repulsion force due to the crowdness:
\begin{equation*}
g(\gamma)=\gamma_+=\left\{
\begin{array}{ll}
    0,  &\text{ if } \gamma\le 0,\\
    \gamma,  &\text{ otherwise}.
\end{array}
\right.
\end{equation*}
In (\ref{sf}b), $\bm{f}_{i}^{\rm wall}$ represents the interaction force of pedestrian $i$ with each wall: 
\begin{align*}
    \bm{f}_{i}^{\rm wall}=\left[A_{i}\exp\left(\frac{r_{i}-d_{i}^{\rm wall}}{B_{i}}\right)+kg\left(r_{i}-d_{i}^{\rm wall}\right)\right]\bm{n}_{i}^{\rm wall}+\kappa g\left(r_{i}-d_{i}^{\rm wall}\right)\langle\bm{v}_{i}, \bm{t}_{i}^{\rm wall}\rangle\bm{t}_{i}^{\rm wall},
\end{align*}
where $d_{i}^{\rm wall}$ measures the distance to the specific wall, $\bm{n}_{i}^{\rm wall}$ the direction perpendicular to it, and $\bm{t}_{i}^{\rm wall}$
the direction tangential to it.


In this paper, we mainly focus on the interaction among individuals. For simplicity, we make further assumptions on the model. 
 Suppose that pedestrians are identical, i.e,  $A_{i}=A$, $B_{i}=B$ and $m_{i}=m$. 
 Furthermore, we assume that the pedestrians are far away from solid walls except the destination or exit, which results in $\bm{f}_{i}^{\rm wall}=\bm{0}$.
 Thus the initial value problem of the social force model which highlights the pairwise interaction can be rewritten as 
\begin{equation}
    \left\{
    \begin{aligned}
    \frac{d\bm{x}_{i}}{dt}&=\bm{v}_{i}, \\
    \frac{d\bm{v}_{i}}{dt}&=\frac{\bm{u}^{\rm e}(\bm{x}_{i},t)-\bm{v}_{i}}{\tau}+\sum_{j\ne i}\Big({\hat{\phi}(\bm{x}_{i},\bm{x}_{j})\bm{n}_{i,j}
    +\hat{\psi}(\bm{x}_{i},\bm{x}_{j})\langle\bm{v}_{j}-\bm{v}_{i}, \bm{t}_{i,j}\rangle\bm{t}_{i,j}}\Big), \\
    \bm{x}_{i}(0) &= \bm{x}_{i,0}, \quad \bm{v}_{i}(0) = \bm{v}_{i,0},
    \end{aligned}
    \right.
    \label{sf1}
\end{equation}
where 
\begin{equation}\label{SF_repulsion}
\small
        \hat{\phi}(\bm{x}_{i},\bm{x}_{j})=\frac{1}{m}\left[A\exp\left(\frac{r_{i,j}-d_{i,j}}{B}\right)+kg(r_{i,j}-d_{i,j})\right],\\
        \hat{\psi}(\bm{x}_{i},\bm{x}_{j})=\frac{1}{m}\kappa g(r_{i,j}-d_{i,j}),
\end{equation}
and $\bm{u}^{\rm e}(\bm{x},t)= U^{\rm e}\big(\bm{x},t\big)\bm{e}\big(\bm{x},t\big)$ is determined by \eqref{eq_desired_speed} and \eqref{eq_desired_direction}.

\subsection{Asymptotic behaviors with a constant desired velocity}

In this section, we consider the special case with a constant desired velocity $\bm{u}^{\rm e}$ and prove the asymptotic behaviors of SFM.
For this scenario, all individuals are aware of the precise location of the single exit. They try to align with the desired velocity, in order to exit the room in a certain optimal way.
Let $M_{j}(t),\ j=0,1$ denote the mass and the moment of velocity:
\begin{align}
    M_{0}(t)=\frac{1}{N}\sum_{i=1}^{N}1=1,\qquad 
    M_{1}(t)=\frac{1}{N}\sum_{i=1}^{N}\bm{v}_{i}(t).
\end{align}
Next, we estimate these moments as follows:

\begin{Pro}
Let $\big(\bm{x}_{i}(t), \bm{v}_{i}(t)\big)$ be the solution to the social force model \eqref{sf1} with $\bm{u}^{\rm e}$ being a constant vector. The following estimates hold:
\begin{equation}
    \begin{aligned}
    \frac{d}{dt}M_{0}(t) &=0,\\
    M_{1}(t)&=\left(M_{1}(0)-\bm{u}^{\rm e}\right)e^{-\frac{t}{\tau}}+\bm{u}^{\rm e}.
    \end{aligned}
    \label{eq_moments_time}
\end{equation}
Hence, the system satisfies the mass conservation, and $M_1(t)$ converges to $\displaystyle \bm{u}^{\rm e}$ exponentially fast as $t\to \infty$.
\label{moment}
\end{Pro}

\begin{proof}

Conservation of mass in equation (\ref{eq_moments_time}) is directly derived since there is no addition or extinction of particles. Next, we consider the first moment.  
Notice that functions $\hat{\phi}(\bm{x}_{i},\bm{x}_{j})$ and $\hat{\psi}(\bm{x}_{i},\bm{x}_{j})$ are symmetric, i.e., 
   \begin{align*}
       \hat{\phi}(\bm{x}_{i},\bm{x}_{j})=\hat{\phi}(\bm{x}_{j},\bm{x}_{i}),\ \ \ \hat{\psi}(\bm{x}_{i},\bm{x}_{j})=\hat{\psi}(\bm{x}_{j},\bm{x}_{i}).
   \end{align*}
   Moreover, we have
   \begin{align*}
       \langle\bm{v}_{i}-\bm{v}_{j}, \bm{t}_{j,i}\rangle = (\bm{v}_{i}-\bm{v}_{j})\cdot \bm{t}_{ji}=(\bm{v}_{j}-\bm{v}_{i})\cdot \bm{t}_{ij}= \langle\bm{v}_{j}-\bm{v}_{i}, \bm{t}_{i,j}\rangle.
   \end{align*}
It follows that
   \begin{equation*}
           \sum_{i=1}^{N}\sum_{j\ne i}\hat{\phi}(\bm{x}_{i},\bm{x}_{j}) \bm{n}_{i,j} = 0, \quad
           \sum_{i=1}^{N}\sum_{j\ne i}\hat{\psi}(\bm{x}_{i},\bm{x}_{j})\langle\bm{v}_{j}-\bm{v}_{i}, \bm{t}_{i,j}\rangle\bm{t}_{i,j}=0.
           \label{sym}
   \end{equation*}
Substituting the above equation into the second equation in (\ref{sf1}), we obtain
    \begin{equation*}
    \begin{aligned}
        \frac{d}{dt}M_{1}(t)&=\frac{1}{N}\sum_{i=1}^{N}\frac{d\bm{v}_{i}}{dt}\\
        &=\frac{1}{N}\sum_{i=1}^{N}\left[\frac{\bm{u}^{\rm e}-\bm{v}_{i}}{\tau}+\sum_{j\ne i}{\hat{\phi}(\bm{x}_{i},\bm{x}_{j})\bm{n}_{i,j}+\hat{\psi}(\bm{x}_{i},\bm{x}_{j})\langle\bm{v}_{j}-\bm{v}_{i}, \bm{t}_{i,j}\rangle\bm{t}_{i,j}}\right]\\
        &=\frac{1}{N}\sum_{i=1}^{N}\frac{\bm{u}^{\rm e}-\bm{v}_{i}}{\tau}
        =\frac{\bm{u}^{\rm e}}{\tau} -\frac{M_{1}(t)}{\tau}.
    \end{aligned}
    \end{equation*}
Solving the above differential equation yields that
\begin{equation*}
\begin{aligned}
    M_{1}(t)&=e^{\int_{0}^{t}-\frac{1}{\tau}dr}\left[M_{1}(0)+\int_{0}^{t}e^{-\int_{0}^{s}-\frac{1}{\tau}dr}\left(\frac{\bm{u}^{\rm e}}{\tau}\right)ds\right]\\
    &=\left(M_{1}(0)-\bm{u}^{\rm e}\right)e^{-\frac{t}{\tau}}+\bm{u}^{\rm e},
    \end{aligned}
\end{equation*}
where $M_{1}(0)$ is the initial moment.
It follows that
\begin{equation*}
    \lim_{t \to \infty} M_{1}(t)=\bm{u}^{\rm e}.
\end{equation*}
\end{proof}

\section{Mean Field models and hydrodynamic Limits}\label{secMeanFieldModels}

We assume that all pedestrians have the same minimum distance from each other, i.e., $r_{i} = {\rm constant}$ for $i = 1, 2, \dots, N$. In this case, $\hat{\phi}$ and $\hat{\psi}$ in (\ref{SF_repulsion}) only depend on the distance between pedestrians $i$ and $j$, and hence we introduce new notations $\Phi$ and $\Psi$ as follows:
\[
\hat{\phi}(\bm{x}_i, \bm{x}_j) = \Phi(\bm{x}_i-\bm{x}_j), \qquad 
\hat{\psi}(\bm{x}_i, \bm{x}_j) = \Psi(\bm{x}_i-\bm{x}_j).
\]
Using the so-called “weak coupling scaling” assumption \cite{braun1977vlasov,neunzert1977vlasov,spohn2012large},
we re-scale the interaction potential with the factor $\frac {1}{N}$. 
Thus, the updated social force model is defined as:
\begin{equation}
\small
    \left\{
    \begin{aligned}
    \frac{d\bm{x}_{i}}{dt}&=\bm{v}_{i}, \\
    \frac{d\bm{v}_{i}}{dt}&= \frac{\bm{u}^{\rm e}(\bm{x}_i,t)-\bm{v}_{i}}{\tau}+\frac{1}{N}\sum_{j\ne i}\Big({\Phi(\bm{x}_{i}-\bm{x}_{j})\bm{n}_{i,j}+\Psi(\bm{x}_{i}-\bm{x}_{j})\langle\bm{v}_{j}-\bm{v}_{i}, \bm{t}_{i,j}\rangle\bm{t}_{i,j}}\Big).
    \end{aligned}
    \right.
    \label{sf2}
\end{equation}
Denote the empirical distribution as $f^{N}(\bm{x},\bm{v},t)$, which is defined by 
\begin{align}
    f^{N}(\bm{x},\bm{v},t)=\frac{1}{N}\sum_{j=1}^{N}\delta\Big(\bm{x}-\bm{x}_{j}(t)\Big)\delta\Big(\bm{v}-\bm{v}_{j}(t)\Big),
\end{align}
where $\delta(\cdot)$ denote the Dirac delta on $\mathbb{R}^{2}$. 
Then send $N\to \infty$, and we formally obtain the mean-field limit for the probability distribution function $f(\bm{x},\bm{v},t)$ of this model:
\begin{equation}\label{mf1}
\begin{aligned}
     &\partial_{t} f+ \bm{v}\cdot \nabla_{\bm{x}}f+\nabla_{\bm{v}}\cdot\left(Ff\right)=0,\\
&F(\bm{x},\bm{v},t)=F_1(\bm{x},\bm{v},t)+F_2(\bm{x},\bm{v},t)+F_3(\bm{x},\bm{v},t),\\
     &F_{1}(\bm{x},\bm{v},t)=\frac{\bm{u}^{\rm e}(\bm{x},t)-\bm{v}}{\tau},\\
     &F_{2}(\bm{x},\bm{v},t)=\int_{\mathbb{R}^{2}}\Phi(\bm{x}-\bm{y})\frac{\bm{x}-\bm{y}}{\Vert \bm{x}-\bm{y}\Vert }\rho(\bm{y},t)d\bm{y},\\
     &F_{3}(\bm{x},\bm{v},t)=\int_{\mathbb{R}^{2}}\frac{\Psi(\bm{x}-\bm{y})}{\Vert \bm{x}-\bm{y} \Vert^{2}}\big\langle\bm{u}(\bm{y},t)-\bm{v},(\bm{x}-\bm{y})^{\perp}\big\rangle(\bm{x}-\bm{y})^{\perp}\rho(\bm{y},t)d\bm{y},\\
     &\rho(\bm{y},t)=\int_{R^2}f(\bm{y},\bm{w},t)d\bm{w},\ \quad \rho(\bm{y},t)\bm{u}(\bm{y},t)=\int_{R^2}\bm{w}f(\bm{y},\bm{w},t)d\bm{w}.
\end{aligned}
\end{equation}
The desired velocity $\bm{u}^{\rm e}(\bm{x},t)$ is determined by the optimal route choice strategy and the corresponding eikonal equation as discussed in Section \ref{secMicroModels}. For the derivation of the above mean field limit, we refer to the relevant literature, for instance, \cite{Sznitman1991}.

\subsection{Hydrodynamic scaling}
The social force model and the above derived mean-field model are defined at the microscopic level, i.e., they describe characteristics of the individual particle at physical time and space scales. Next, we aim to study dynamics of the system at large time and space scales. Hence, the following change of variables are introduced:
\begin{align}
\tilde{\bm{x}}=\epsilon \bm{x},\ \ \tilde t=\epsilon t\quad \text{ for  }\quad  0< \epsilon \ll 1,
\end{align}
We also denote $\tilde \tau=\epsilon\tau$ as the scaled intensity. For notational simplicity, we will omit the tildes hereafter, and accordingly, the symbols $\bm{x}$, $t$, and $\tau$ denote the new scaled variables. Based on (\ref{mf1}), we can derive that the scaled distribution function $f^{\epsilon}$ satisfies:
\begin{equation}
    \begin{aligned}
    &\epsilon\left(\partial_{t} f^{\epsilon}+ \bm{v}\cdot \nabla_{\bm{x}}f^{\epsilon}\right)+\nabla_{\bm{v}}\cdot\left(F^{\epsilon}f^{\epsilon}\right)=0,\\
    &F^{\epsilon}_{1}(\bm{x},\bm{v},t)=\epsilon\frac{\bm{u}^{\rm e}(\bm{x},t)-\bm{v}}{\tau},\\
     &F^{\epsilon}_{2}(\bm{x},\bm{v},t)=\int_{\mathbb{R}^{2}}\Phi\left(\frac{\bm{x}-\bm{y}}{\epsilon}\right)\frac{\bm{x}-\bm{y}}{\Vert \bm{x}-\bm{y}\Vert }\rho^{\epsilon}(\bm{y},t)d\bm{y},\\
     &F^{\epsilon}_{3}(\bm{x},\bm{v},t)=\int_{\mathbb{R}^{2}}\frac{\Psi\left(\frac{\bm{x}-\bm{y}}{\epsilon}\right)}{\Vert \bm{x}-\bm{y}\Vert^{2}}\langle\bm{u}(\bm{y},t)-\bm{v},(\bm{x}-\bm{y})^{\perp}\rangle(\bm{x}-\bm{y})^{\perp}\rho^{\epsilon}(\bm{y},t)d\bm{y}, 
    \end{aligned}
\end{equation}
where the average density and flux are given by 
\[
     \rho^{\epsilon}(\bm{y},t)=\int_{\mathbb{R}^{2}}f^{\epsilon}(\bm{y},\bm{w},t)d\bm{w},\qquad \rho^{\epsilon}(\bm{y},t)\bm{u}^{\epsilon}(\bm{y},t)=\int_{\mathbb{R}^{2}}\bm{w}f^{\epsilon}(\bm{y},\bm{w},t)d\bm{w}.
\]

\subsection{The interaction kernels}


Besides the specific choice in  \eqref{SF_repulsion}, we now discuss general particle interaction functions $\phi(r)$ and $\psi(r)$ that are of order 1, nonnegative, monotonically decreasing, and satisfy the integratibility condition with
\[
\gamma_1 = \pi\int_{0}^{\infty}\phi(r)r^2\,dr < \infty, \quad
\gamma_2 = \pi\int_{0}^{\infty}\psi(r)r\,dr < \infty.
\]
Here $r=\|\bm{x}-\bm{y}\|$ is the distance between two pedestrians located at $\bm{x}$ and $\bm{y}$. We do not distinguish $\phi(\bm{x}-\bm{y})$ and $\phi(\|\bm{x}-\bm{y}\|)$ when there is no ambiguity.
According to the different ranges and intensities of $\Phi$ and $\Psi$,  the forces $F_{2}^{\epsilon}$ and $F_{3}^{\epsilon}$ have different corresponding expressions.

\begin{figure}[htbp]
	\centering
	\subfigure[$\Phi(\bm{x})$]{
		\includegraphics[width=2.in]{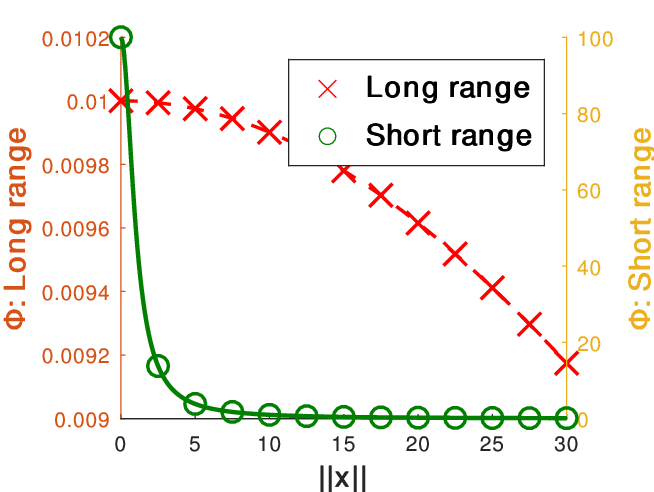} \label{fig_Phi}
	}
	\subfigure[$\Psi(\bm{x})$]{
		\includegraphics[width=2.in]{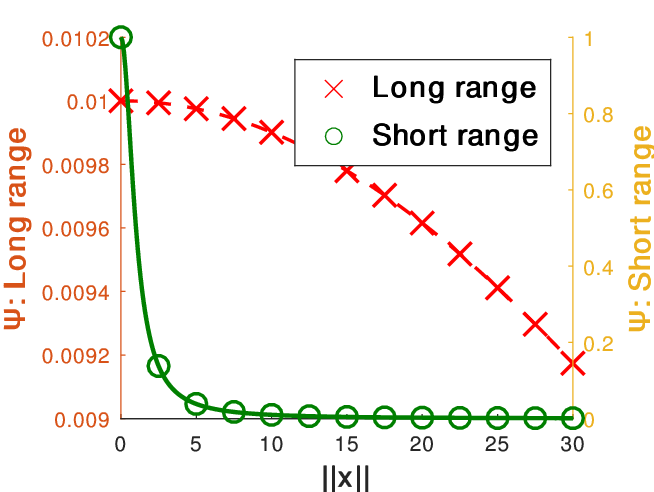} \label{fig_Psi}
	}
	\caption{The interaction kernels $\Phi(\bm{x})$ and $\Psi(\bm{x})$ with different range and $\epsilon=0.01$.} \label{fig_interaction_kernel}
\end{figure}

\textbf{Case 1:} The repulsion force between particles are weak in the sense that
\[
\Phi\left(\frac{\bm{x}-\bm{y}}{\epsilon}\right)=\epsilon\phi(\bm{x}-\bm{y}), \quad \text{ and } \quad
\Psi\left(\frac{\bm{x}-\bm{y}}{\epsilon}\right)=\epsilon \psi(\bm{x}-\bm{y}).
\]
This indicates that the particle system has a long-range psychological force and body force, and the intensity strength is scaled by $\epsilon$.
Thus these forces remain non-local as $\epsilon$ tends to 0. 
Meanwhile, the particle system has a long-range sliding friction force and the intensity is scaled by $\epsilon$ as well. 
FIG. \ref{fig_interaction_kernel}(a) and \ref{fig_interaction_kernel}(b) with left vertical axis and red crosses show that the interaction strength has a long large tail as $\|x\|$ increases. 
For this illustration, we take $\epsilon = 0.01$ and the Cucker-Smale type interaction kernel:
\[\phi(\bm{x}) = \psi(\bm{x}) = \frac{1}{1+\|\bm{x}\|^2}.\]
So we have 
    \begin{equation}
    \begin{aligned}
    F^{\rm {w}, \epsilon}_{2}(\bm{x},\bm{v},t)&:=\epsilon\int_{\mathbb{R}^{2}}\phi(\bm{x}-\bm{y})\frac{\bm{x}-\bm{y}}{\Vert \bm{x}-\bm{y}\Vert }\rho^{\epsilon}(\bm{y},t)d\bm{y},\\
    F^{\rm w, \epsilon}_{3}(\bm{x},\bm{v},t)
        &:=\epsilon\int_{\mathbb{R}^{2}}\frac{\psi(\bm{x}-\bm{y})}{\Vert \bm{x}-\bm{y}\Vert^{2}}\langle\bm{u}^{\epsilon}(\bm{y},t)-\bm{v},(\bm{x}-\bm{y})^{\perp}\rangle(\bm{x}-\bm{y})^{\perp}\rho^{\epsilon}(\bm{y},t)d\bm{y}.
        \end{aligned}
    \end{equation}


\textbf{Case 2:} The repulsion force between particles are strong in the sense that
\[  
\Phi\left(\frac{\bm{x}-\bm{y}}{\epsilon}\right)=\frac{1}{\epsilon}\phi\left(\frac{\bm{x}-\bm{y}}{\epsilon}\right),
\quad \text{ and } \quad
\Psi\left(\frac{\bm{x}-\bm{y}}{\epsilon}\right)= \psi\left(\frac{\bm{x}-\bm{y}}{\epsilon}\right) .
\]
This indicates that the particle system has a short-range psychological force and body force, and the intensity is scaled by $1/\epsilon$. Thus the range of these forces are supposed to be of order $1/\epsilon$.
This indicates that the particle system has a short-range sliding friction force and the magnitude of the intensity remains unchanged. 
FIG. \ref{fig_interaction_kernel}(a) and \ref{fig_interaction_kernel}(b) with right vertical axis and green circles show that the interaction strength is strong when the distance between individuals is short, and decreases dramatically as $\|x\|$ increases. 
It follows that
\begin{equation}
	\begin{aligned}
		F^{\epsilon}_{2}(\bm{x},\bm{v},t)&=\frac{1}{\epsilon}\int_{\mathbb{R}^{2}}\phi\left(\frac{\bm{x}-\bm{y}}{\epsilon}\right)\frac{\bm{x}-\bm{y}}{\Vert \bm{x}-\bm{y}\Vert }\rho^{\epsilon}(\bm{y},t)d\bm{y}\\
		&=\frac{1}{\epsilon}\int_{\mathbb{R}^{2}}\phi(\bm{z})\frac{\bm{z}}{\Vert \bm{z}\Vert }\rho^{\epsilon}(\bm{x}-\epsilon \bm{z},t)(-\epsilon)d\bm{z}\\
		&=-\int_{\mathbb{R}^{2}}\phi(\bm{z})\frac{\bm{z}}{\Vert \bm{z}\Vert }\left[\rho^{\epsilon}(\bm{x},t)-(\epsilon \bm{z})\nabla_{\bm{x}}\rho^{\epsilon}(\bm{x},t)+O(\epsilon^{2})\right]d\bm{z}.
	\end{aligned}
\end{equation}
where $\bm{z}=\frac{\bm{x}-\bm{y}}{\epsilon}$. 
Since $\phi(\bm{z})$ is an even function, we have
\begin{align}
	\int_{\mathbb{R}^{2}}\phi(\bm{z})\frac{\bm{z}}{\Vert \bm{z}\Vert }\rho^{\epsilon}(\bm{x},t)d\bm{z} =0.
\end{align}
Thus
\begin{equation}
	\begin{aligned}
		F^{\epsilon}_{2}(\bm{x},\bm{v},t)= F^{\rm s,\epsilon}_{2}(\bm{x},\bm{v},t)+O(\epsilon^{2})
		\quad \text{ with }\quad
		F^{\rm s,\epsilon}_{2}(\bm{x},\bm{v},t):=\epsilon \gamma_1\nabla_{\bm{x}}\rho^{\epsilon}(\bm{x},t). 
	\end{aligned}
\end{equation}
Moreover, we have
\begin{equation}
	\begin{aligned}
		F^{\epsilon}_{3}(\bm{x},\bm{v},t)&=\int_{\mathbb{R}^{2}}\frac{\Psi\left(\frac{\bm{x}-\bm{y}}{\epsilon}\right)} {\Vert \bm{x}-\bm{y}\Vert^{2}}\langle\bm{u}^{\epsilon}(\bm{y},t)-\bm{v},(\bm{x}-\bm{y})^{\perp}\rangle(\bm{x}-\bm{y})^{\perp}\rho^{\epsilon}(\bm{y},t)d\bm{y}\\
		&=\int_{\mathbb{R}^{2}}\frac{\psi\left(\frac{\bm{x}-\bm{y}}{\epsilon}\right)} {\Vert \bm{x}-\bm{y}\Vert^{2}}\langle\bm{u}^{\epsilon}(\bm{y},t)-\bm{v},(\bm{x}-\bm{y})^{\perp}\rangle(\bm{x}-\bm{y})^{\perp}\rho^{\epsilon}(\bm{y},t)d\bm{y}\\
		&=-\epsilon\int_{\mathbb{R}^{2}}\frac{\psi(\bm{z})} {\Vert \bm{z}\Vert^{2}}\langle\bm{u}^{\epsilon}(\bm{x}-\epsilon \bm{z},t)-\bm{v},\bm{z}^{\perp}\rangle\bm{z}^{\perp}\rho^{\epsilon}(\bm{x}-\epsilon \bm{z},t)d\bm{z}\\
		&=-\epsilon\int_{\mathbb{R}^{2}}\frac{\psi(\bm{z})} {\Vert \bm{z}\Vert^{2}}\langle\bm{u}^{\epsilon}(\bm{x},t)-\nabla_{\bm{x}}\bm{u}^{\epsilon}(\bm{x},t)(\epsilon \bm{z})+O(\epsilon^{2})-\bm{v},\bm{z}^{\perp}\rangle\bm{z}^{\perp}\\
        &\qquad\qquad \qquad \qquad \left[\rho^{\epsilon}(\bm{x},t)-(\epsilon \bm{z})\nabla_{\bm{x}}\rho^{\epsilon}(\bm{x},t)+O(\epsilon^{2})\right]d\bm{z}\\
		&=-\epsilon\int_{\mathbb{R}^{2}}\frac{\psi(\bm{z})}{\Vert \bm{z}\Vert^{2}}\left( \rho^{\epsilon}(\bm{x},t)\langle\bm{u}^{\epsilon}(\bm{x},t)-\bm{v}, \bm{z}^{\perp}\rangle+O(\epsilon)\right)\bm{z}^{\perp}d\bm{z}\\
		&=-\epsilon \rho^{\epsilon}(\bm{x},t)\int_{\mathbb{R}^{2}}\frac{\psi(\bm{z})}{\Vert \bm{z}\Vert^{2}}\bm{z}^{\perp}\otimes \bm{z}^{\perp}d\bm{z}(\bm{u}^{\epsilon}(x,t)-\bm{v})+O(\epsilon^{2})\\
		&=-\epsilon \gamma_2\rho^{\epsilon}(\bm{x},t)\left(\bm{u}^{\epsilon}(\bm{x},t)-\bm{v}\right)+O(\epsilon^{2})=:F_3^{\rm s, \epsilon}(\bm{x},\bm{v},t) + O(\epsilon^{2}).
	\end{aligned}
\end{equation}


Based on above analysis, letting $\epsilon \to 0$,  we obtain 
\begin{equation}
    \partial_{t} f+ \bm{v}\cdot \nabla_{\bm{x}}f+\nabla_{\bm{v}}\cdot\left(Ff\right)=0,
    \label{mf2}
\end{equation}
where $F=F_{1}+F_{2}+F_{3}$ and $F_{1}=\frac{\bm{u}^{\rm e}(\bm{x},t)-\bm{v}}{\tau}$. In the weak case, $F_2$ and $F_3$ become
\begin{equation}
    \begin{aligned}
        F_{2}^{\rm w}&=\int_{\mathbb{R}^{2}}\phi(\bm{x}-\bm{y})\frac{\bm{x}-\bm{y}}{\Vert \bm{x}-\bm{y}\Vert }\rho(\bm{y},t)d\bm{y},\\
        F_{3}^{\rm w}&=\int_{\mathbb{R}^{2}}\frac{\psi(\bm{x}-\bm{y})}{\Vert \bm{x}-\bm{y}\Vert^{2}}\langle\bm{u}(\bm{y},t)-\bm{v},(\bm{x}-\bm{y})^{\perp}\rangle(\bm{x}-\bm{y})^{\perp}\rho(\bm{y},t)d\bm{y}.
    \end{aligned}
\end{equation}
In the strong case, we take $F_2$ and $F_3$ as
\begin{equation}
    \begin{aligned}
F_{2}^{\rm s}&=\gamma_{1}\nabla_{\bm{x}}\rho(\bm{x},t), \\
F_{3}^{\rm s}&=\gamma_{2}\rho(\bm{x},t)(\bm{u}(\bm{x},t)-\bm{v}).
    \end{aligned}
\end{equation}

\section{Macroscopic model}\label{secMacroModels}
In this subsection, we establish the macroscopic model based on the above microscopic SFM and mesoscopic model. 
In other words, we will derive the system for the density $\rho(\bm{x},t)$ and the first moment $(\rho \bm{u})(\bm{x},t)$. 
\begin{Thm}
The macroscopic model for the kinetic system \eqref{mf1} is given by
\begin{equation}
\small
    \begin{aligned}
        \partial_{t} \rho &+\nabla_{\bm{x}}\cdot(\rho \bm{u})=0,\\
        \partial_{t} (\rho \bm{u}) &+\nabla_{x}\cdot(\rho \bm{u}\otimes \bm{u})=\rho\frac{\bm{u}^{\rm e}(\bm{x},t)-\bm{u}}{\tau} 
        +\rho(\bm{x},t)\int_{\mathbb{R}^{2}}\Phi(\bm{x}-\bm{y})\frac{\bm{x}-\bm{y}}{\Vert \bm{x}-\bm{y}\Vert }\rho(\bm{y},t)d\bm{y}\\
        &+\rho(\bm{x},t)\int_{\mathbb{R}^{2}}\frac{\Psi(\bm{x}-\bm{y})}{\Vert \bm{x}-\bm{y}\Vert^{2}}\langle\bm{u}(\bm{y},t)-\bm{u}(\bm{x},t),(\bm{x}-\bm{y})^{\perp}\rangle(\bm{x}-\bm{y})^{\perp}\rho(\bm{y},t) d\bm{y}.
        \label{mac1}
    \end{aligned}
\end{equation}
\end{Thm}

\begin{proof}
For kinetic equation (\ref{mf1}), take the test function $\zeta=1$. Then integration about the variable $v$ leads to 
\begin{equation}
    \begin{aligned}
        &\int_{\mathbb{R}^{2}} \big[\partial_{t} f+ \bm{v}\cdot \nabla_{\bm{x}}f+\nabla_{\bm{v}}(Ff)\big]d\bm{v}=0.\\
        \Rightarrow \;&\partial_{t}\int_{\mathbb{R}^{2}}fd\bm{v}+\nabla_{\bm{x}}\cdot\int_{\mathbb{R}^{2}} f\bm{v}d\bm{v} =0.\\
        \Rightarrow \;&\partial_{t} \rho +\nabla_{\bm{x}}\cdot(\rho \bm{u})=0.
    \end{aligned}
\end{equation}
Take the test function $\zeta=v$, and then we have
\begin{equation}
    \begin{aligned}
        &\int_{\mathbb{R}^{2}} \big[\partial_{t} f+ \bm{v}\cdot \nabla_{\bm{x}}f+\nabla_{\bm{v}}(Ff)\big]\bm{v}d\bm{v}=0.\\
        \Rightarrow \;&\partial_{t} \int_{\mathbb{R}^{2}} f\bm{v}d\bm{v}+\nabla_{\bm{x}}\cdot\int_{\mathbb{R}^{2}} f(\bm{v}\otimes \bm{v})d\bm{v}+\int_{\mathbb{R}^{2}}\nabla_{\bm{v}}\cdot (Ff)\bm{v}d\bm{v}=0.
    \end{aligned}
\end{equation}
Here the second and third terms satisfy
\begin{equation}
    \begin{aligned}
        &\int_{\mathbb{R}^{2}} f(\bm{v}\otimes \bm{v})d\bm{v}=\int_{\mathbb{R}^{2}} f(\bm{u}+\bm{v}-\bm{u})\otimes (\bm{u}+\bm{v}-\bm{u})d\bm{v}\\
        &=\int_{\mathbb{R}^{2}} f(\bm{u}\otimes \bm{u})d\bm{v}+\int_{\mathbb{R}^{2}} f(\bm{u}-\bm{v})\otimes (\bm{u}-\bm{v})d\bm{v}=\rho \bm{u}\otimes \bm{u} +P.
    \end{aligned}
\end{equation}
\begin{equation}
    \begin{aligned}
        \int_{\mathbb{R}^{2}}\nabla_{\bm{v}}\cdot (Ff)\bm{v}d\bm{v}=Ff|_{\partial \mathbb{R}^{2}} -\int_{\mathbb{R}^{2}} Ff d\bm{v}=-\int_{\mathbb{R}^{2}} Ff d\bm{v}.
    \end{aligned}
\end{equation}
According to the definition of $F$, we have:
\begin{align}
    \int_{\mathbb{R}^{2}} F_{1}f d\bm{v}=\int_{\mathbb{R}^{2}} \frac{\bm{u}^{\rm e}-\bm{v}}{\tau}f d\bm{v}=\rho\frac{\bm{u}^{\rm e}-\bm{u}}{\tau}.
\end{align}
To close the system, we assume the separation of variables for $f$ in the sense that 
\begin{align}
    f(\bm{x},\bm{v},t)=\rho(\bm{x},t)\delta\left(\bm{u}(\bm{x},t)-\bm{v}\right).
\end{align}
So $P=0$.
Thus based on original kinetic equation (\ref{mf1}), the following direct continuum equation is derived
\begin{equation}
    \begin{aligned}
        \partial_{t} \rho &+\nabla_{\bm{x}}\cdot(\rho \bm{u})=0,\\
        \partial_{t} (\rho \bm{u}) &+\nabla_{x}\cdot(\rho \bm{u}\otimes \bm{u})=\rho\frac{\bm{u}^{\rm e}-\bm{u}}{\tau} 
        +\rho(\bm{x},t)\int_{\mathbb{R}^{2}}\Phi(\bm{x}-\bm{y})\frac{\bm{x}-\bm{y}}{\Vert \bm{x}-\bm{y}\Vert }\rho(\bm{y},t)d\bm{y}\\
        &+\rho(\bm{x},t)\int_{\mathbb{R}^{2}}\frac{\Psi(\bm{x}-\bm{y})}{\Vert \bm{x}-\bm{y}\Vert^{2}}\langle\bm{u}(\bm{y},t)-\bm{u}(\bm{x},t),(\bm{x}-\bm{y})^{\perp}\rangle(\bm{x}-\bm{y})^{\perp}\rho(\bm{y},t) d\bm{y}.
    \end{aligned}
\end{equation}
\end{proof}

\begin{Thm}
    The macroscopic model for the kinetic system \eqref{mf2} with hydrodynamic scaling is given by
\begin{equation}
    \begin{aligned}
        \partial_{t} \rho +\nabla_{\bm{x}}\cdot(\rho \bm{u})&=0,\\
        \partial_{t} (\rho \bm{u}) +\nabla_{\bm{x}}\cdot(\rho \bm{u}\otimes \bm{u})&=\rho\frac{\bm{u}^{\rm e}-\bm{u}}{\tau} +S^{1}+S^{2},
        \label{mac2}
    \end{aligned}
\end{equation}
where $S^{1}$ takes equation (\ref{f21}) or (\ref{f22}), $S^{2}$ takes equation (\ref{f31}) or (\ref{f32}).    
\end{Thm}
\begin{proof}
According to the analysis in Section \ref{secMeanFieldModels}, we have the following two cases:

\textbf{Case 1 (weak case):}  
$$F_{2}=\int_{\mathbb{R}^{2}}\int_{\mathbb{R}^{2}}\phi(\bm{x}-\bm{y})\frac{\bm{x}-\bm{y}}{\Vert \bm{x}-\bm{y}\Vert }\rho(\bm{y},t)d\bm{y}.$$
$$F_{3}=\int_{\mathbb{R}^{2}}\frac{\psi(\bm{x}-\bm{y})}{\Vert \bm{x}-\bm{y}\Vert^{2}}\langle\bm{u}(\bm{y},t)-\bm{v},(\bm{x}-\bm{y})^{\perp}\rangle(\bm{x}-\bm{y})^{\perp}\rho(\bm{y},t)d\bm{y}.$$

In this case, denoting $\bm{z}=\bm{x}-\bm{y}$, we have
    \begin{align}
        \int_{\mathbb{R}^{2}} F_{2}f d\bm{v}&=\int_{\mathbb{R}^{2}}\int_{\mathbb{R}^{2}}\phi(\bm{z})\frac{\bm{z}}{\Vert \bm{z}\Vert }\rho(\bm{y},t)f(\bm{x},\bm{v},t) d\bm{v}d\bm{y}\nonumber\\
       &= \int_{\mathbb{R}^{2}}\phi(\bm{z})\frac{\bm{z}}{\Vert \bm{z}\Vert }\rho(\bm{x},t)\rho(\bm{y},t) d\bm{y}. \label{f21}\\
        \int_{\mathbb{R}^{2}} F_{3}f d\bm{v}&=\int_{\mathbb{R}^{2}}\int_{\mathbb{R}^{2}}\frac{\psi(\bm{z})}{\Vert \bm{z}\Vert^{2}}\langle\bm{u}(\bm{y},t)-\bm{v},\bm{z}^{\perp}\rangle\bm{z}^{\perp}\rho(\bm{y},t)f(\bm{x},\bm{v},t) d\bm{v}d\bm{y}\nonumber\\
       &=  \int_{\mathbb{R}^{2}}\frac{\psi(\bm{z})}{\Vert\bm{z}\Vert^{2}}\langle\bm{u}(\bm{y},t)-\bm{u}(\bm{x},t),\bm{z}^{\perp}\rangle\bm{z}^{\perp}\rho(\bm{x},t)\rho(\bm{y},t) d\bm{y}. 
       \label{f31}
    \end{align}

\textbf{Case 2 (strong case):} $F_{2}=\gamma_{1}\nabla_{\bm{x}}\rho(\bm{x},t)$ and $F_{3}=\gamma_{2}\rho(\bm{x},t)(\bm{u}(\bm{x},t)-\bm{v})$. 

Then we have
\begin{align}
    \int_{\mathbb{R}^{2}} F_{2}f d\bm{v}= \gamma_{1}\int_{\mathbb{R}^{2}}\nabla_{\bm{x}}\rho(\bm{x},t)f d\bm{v}=\gamma_{1}\rho(\bm{x},t)\nabla_{\bm{x}}\rho(\bm{x},t).
    \label{f22}\\
    \int_{\mathbb{R}^{2}} F_{3}f d\bm{v}=\gamma_{2}\int_{\mathbb{R}^{2}} \rho(\bm{x},t)(\bm{u}(\bm{x},t)-\bm{v})f(\bm{x},\bm{v},t) dv=0.
    \label{f32}
\end{align}
Based on the  kinetic equation (\ref{mf2}) with hydrodynamic scaling, the following hydrodynamic continuum equation is derived
\begin{equation*}
    \begin{aligned}
        \partial_{t} \rho +\nabla_{\bm{x}}\cdot(\rho \bm{u})&=0,\\
        \partial_{t} (\rho \bm{u}) +\nabla_{\bm{x}}\cdot(\rho \bm{u}\otimes \bm{u})&=\rho\frac{\bm{u}^{\rm e}-\bm{u}}{\tau} +S^{1}+S^{2},
    \end{aligned}
\end{equation*}
where $S^{1}$ takes equation (\ref{f21}) or (\ref{f22}), $S^{2}$ takes equation (\ref{f31}) or (\ref{f32}).
\end{proof}

\begin{Remark}
For $F_2$ and $F_3$ in \textbf{\rm{Case 1}}, these forces are long-range, and one can derive a nonlocal term in the continuum equation. 
In \textbf{\rm{Case 2}}, these forces are short-range, and one can derive a local term in the continuum equation.
\end{Remark}

\section{Numerical examples}\label{secNumericalExamples}

In this section, numerical examples are established to compare the social force model (\ref{sf1}) and the related direct macroscopic model (\ref{mac1}). The third-order Runge-Kutta method is used to solve the social force model. The efficient fifth-order WENO scheme
in space together with the third-order TVD Runge-Kutta time discretization are used for solving the continuum equation. The fast sweeping method is adopted for the eikonal equation. 

\begin{figure}[htbp]
	\centering
	\subfigure[Example 1]{
		\includegraphics[width=1.54in]{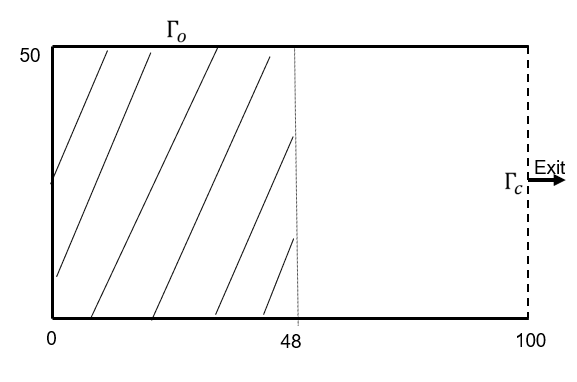} 
	}
	\subfigure[Example 2]{
		\includegraphics[width=1.54in]{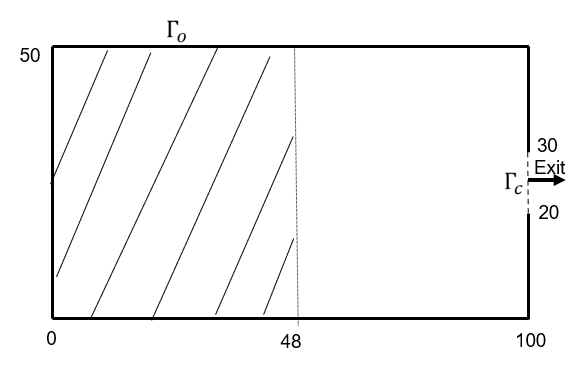} 
	}
	\subfigure[Example 3]{
		\includegraphics[width=1.54in]{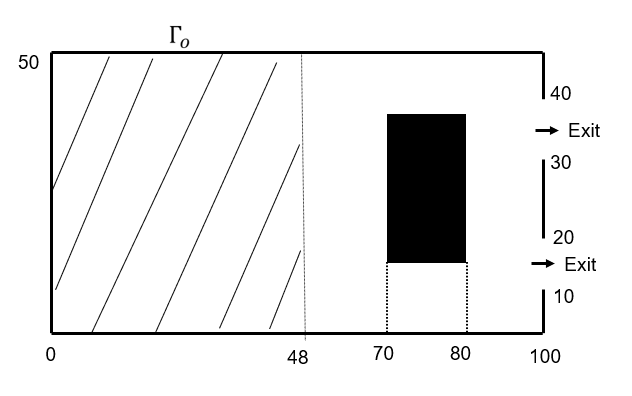} 
	}
	\caption{Three different modeling domains.} \label{fig_region}
\end{figure}

We consider a rectangular room, denoted by $\Omega = L_1 \times L_2 = [0\ {\rm{m}},100\  {\rm{m}}]\times [0 {\rm{m}},50\ {\rm{m}}]$. The exit on the right boundary 
is denoted as $\Gamma_c$, and $\Gamma_o$ is the remaining solid boundary of the room. As shown in FIG \ref{fig_region}, three examples with different exits and obstacles are considered. The first two examples consider obstacle-free cases with different exits: 
\begin{enumerate}
    \item 
    In Example 1, we set $\Gamma_c=\{100\ {\rm{m}}\}\times[0\ {\rm{m}},100\ {\rm{m}}]$, corresponding to a low-density evacuation scenario.
    \item
    In Example 2, we let $\Gamma_c=\{100\ {\rm{m}}\}\times[20\ {\rm{m}},30\ {\rm{m}}]$, which means the exit size is reduced and the narrower passage may lead to a high-density scenario. 
    \item
    Example 3 incorporates an obstacle within the domain, located at $[70\ {\rm{m}}, 80\ {\rm{m}}]$ $ \times [15\ {\rm{m}}, 35\ {\rm{m}}]$. We assume that the pedestrian cannot enter or leave the obstacle. In this example, we set two exits and let $\Gamma_c=\{100\ {\rm{m}}\}\times\{ [10\ {\rm{m}}, 20\ {\rm{m}}] \cup [30\ {\rm{m}}, 40\ {\rm{m}}]\}$.
\end{enumerate}

Initially, there are $N=2400$ pedestrians uniformly distributed in the left part of the room, denoted as $\Omega_{\rm{ini}}=[0\ {\rm{m}},48\ {\rm{m}}]\times [0\ {\rm{m}},50\ {\rm{m}}]$. Then pedestrians move towards the exit $\Gamma_c$ with an identical desired velocity $\bm{u}^{\rm e}$. As to the initial condition setting in the microscopic model, the distance between pedestrians is set as $1$ m such that they do not interact with each other. 
Thus, pedestrians are located at grid points $\bm{x}_{50i+j}:=(i-0.5, j-0.5),\ i=1,...,48,\ j=1,...,50$, where $i$ is the horizontal index and $j$ is the vertical index of the pedestrian.  
Meanwhile, the initial density in the macroscopic model is set as $\rho_{0}(\bm{x})=\chi_{\Omega_{\rm{ini}}}(\bm{x})$ where $\chi$ is the characteristic function. 
When computing the desired direction of the pedestrian by the eikonal equation, the local travel cost is defined as $c(\bm{x},t)=1/U^{\rm e}(\bm{x})$.  
The other values of the parameters and functions are given in Table \ref{tab:my_label} \cite{liang2021continuum,helbing2000simulating}. 

\begin{table}[h!]
	\centering
	\begin{tabular}{ccc}\hline
		Parameters/Functions&  Denotation& Value\\ \hline
		The mass of pedestrian&  $m$& 60\\
		Relaxation time of the  pedestrian&  $\tau$& 0.5 \\ 
		Model parameter&  $\beta$& 0.05\\
		Free-flow speed&  $U_{f}$& 1.034\\
		Measurement parameter&  R& 0.7\\
		The strength of interaction force&  $A_{i}$&  2000 \\
		The range of interaction force&  $B_{i}$& 0.08\\
		The radii of pedestrian $i$&  $r_{i}$& 0.15\\
		The contact strength&  $k$& $1.2\times 10^5$\\
		The friction coefficient&  $\kappa$& $2.4\times 10^5$\\ \hline
	\end{tabular}
	\caption{Parameters and functions.}
	\label{tab:my_label}
\end{table}

The numerical boundary conditions are summarized as follows:
\begin{enumerate}
    \item  In the social force model, to avoid unnecessary calculation, if a pedestrian walks out of the room through the exit, it will be removed from the computational domain.
    \item  On the solid wall boundaries $\Gamma_o$, i.e., the wall of the room except for the exit, to avoid pedestrians walking into the boundary,  we use the reflecting boundary conditions in both the microscopic social force model and the macroscopic continuum model. Specifically, in the social force model, when a pedestrian intersects with these solid boundaries in a time step, we will reflect the displacement and the velocity. In the macroscopic continuum model, we let the normal numerical flux be zero.
\end{enumerate}

\subsection{Comparison between microscopic and macroscopic models}
\begin{figure}[htbp]
	\centering
	\subfigure[Example 1]{
		\includegraphics[width=1.58in]{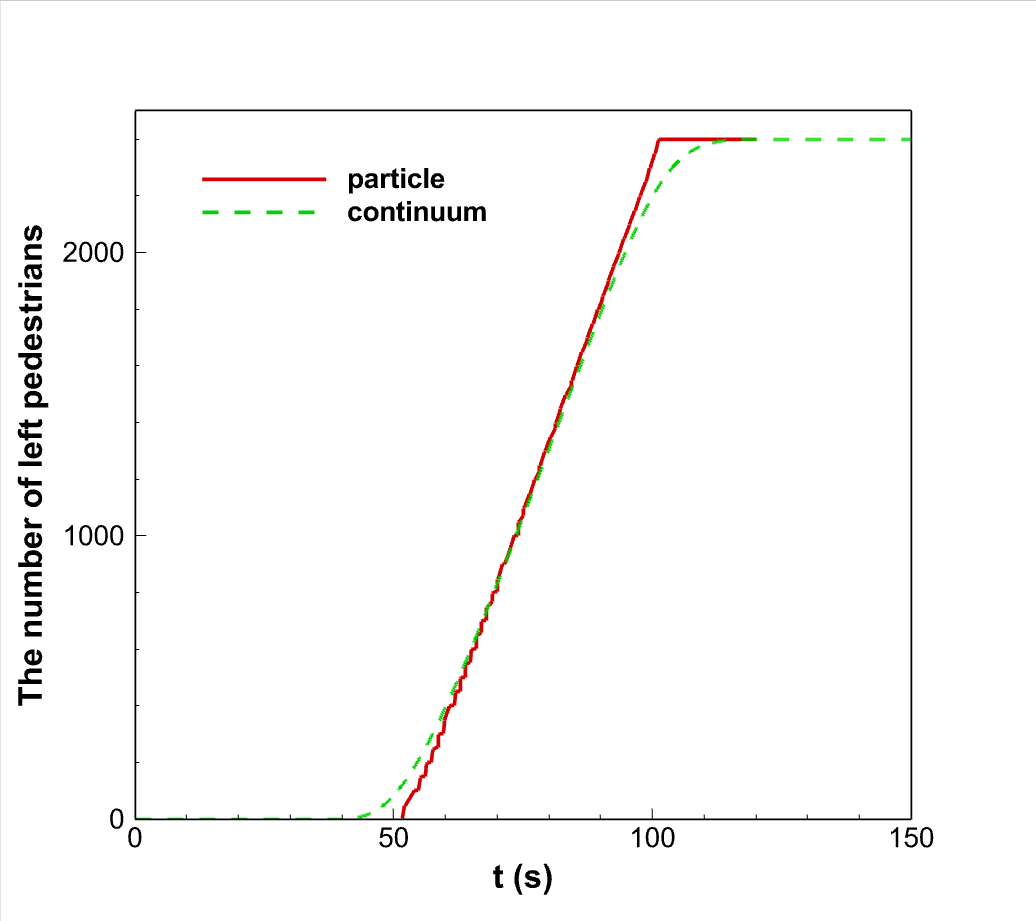} 
	}
	\subfigure[Example 2]{
		\includegraphics[width=1.58in]{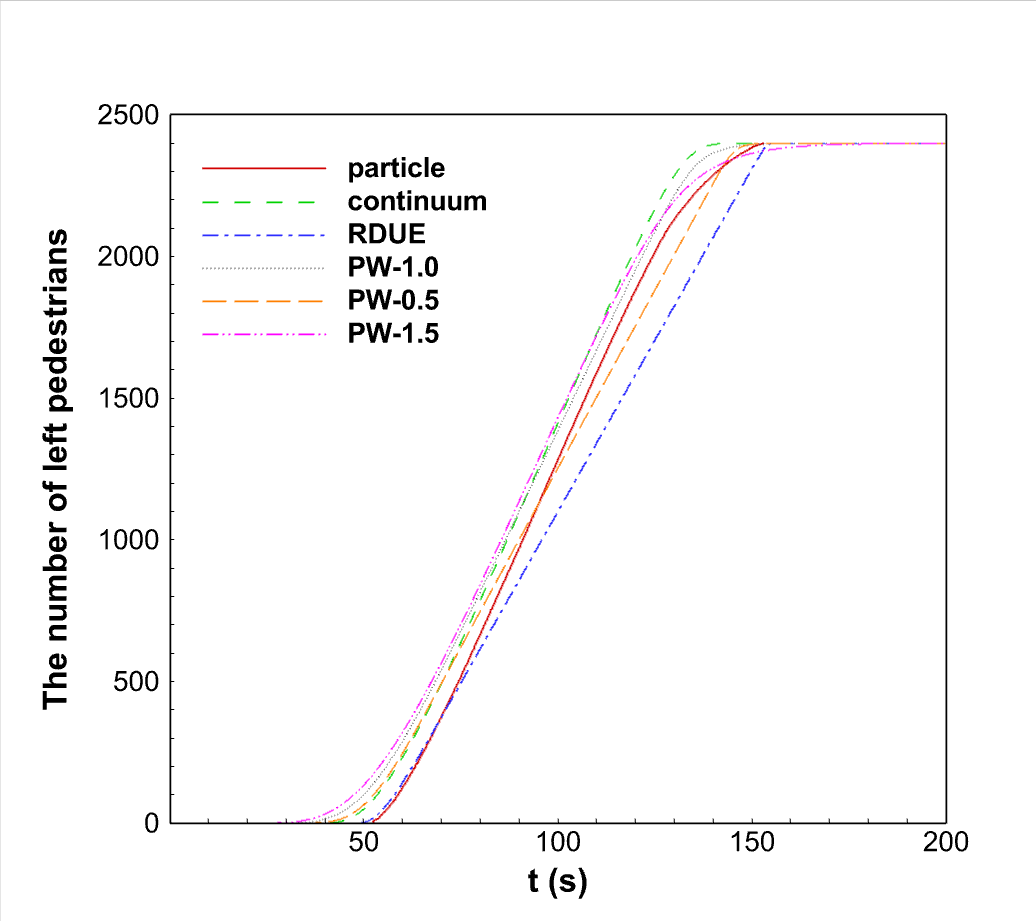} 
	}
	\subfigure[Example 3]{
		\includegraphics[width=1.58in]{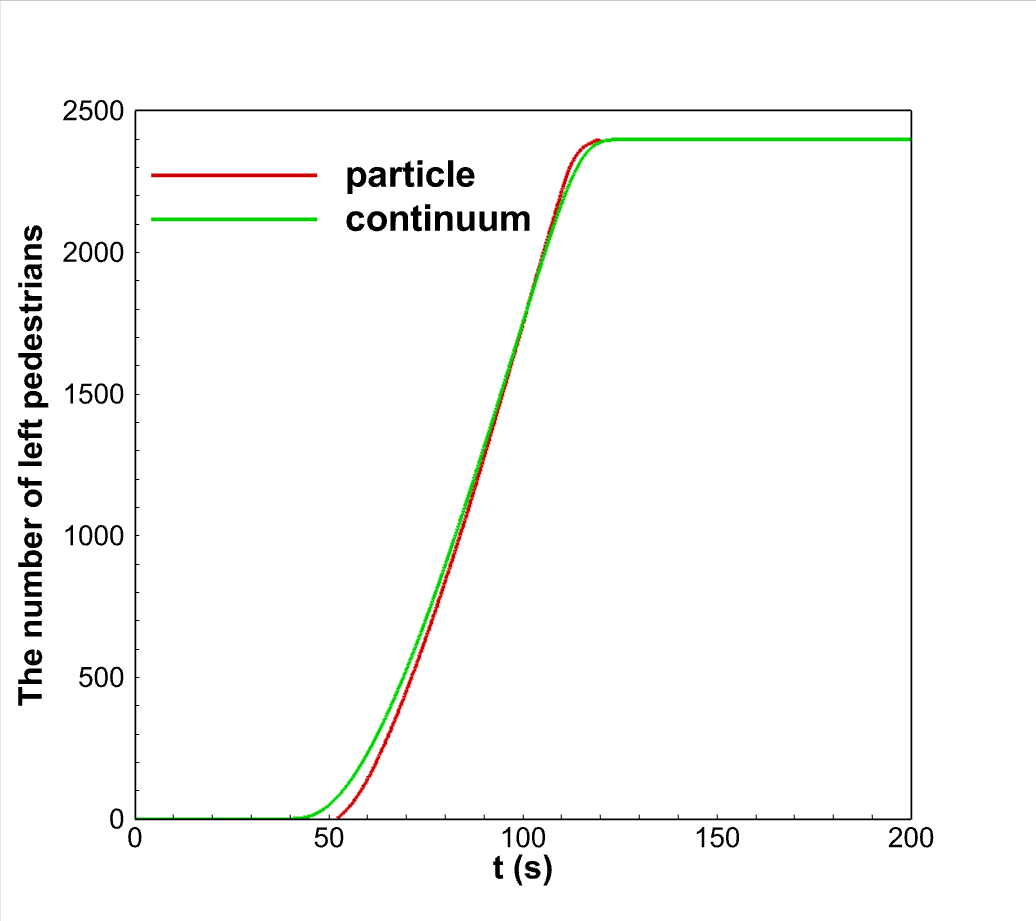} 
	}
	\caption{The number of left pedestrians against the simulation time $t$.} \label{f4}
\end{figure}

In this section, we do some comparisons between microscopic and macroscopic model. FIG. \ref{f4} shows the number of pedestrians who have left the room against the simulation time $t$ in different examples. In sub-figures \ref{f4} (a) and (c), we compare the behavior of the particle model and the related continuum model, in sub-figure \ref{f4} (b), the above particle model, the related continuum model, the reactive dynamic user-equilibrium (RDUE) model \cite{huang2009revisiting}, and the Payne-Whitham (PW) model \cite{payne1971mathematical} with different sonic speed are compared. It can be observed that pedestrians in Example 1 evacuate the room the fastest due to the widest exit. Although an obstacle is present, the larger exit in Example 3 results in the second-shortest evacuation time, while the Example 2 requires the longest time to complete evacuation. These sub-figures also indicate that the particle model and the continuum model yield comparable results in simulating pedestrian evacuation under low-density conditions, verifying the correctness of our derivation. Especially, all models in sub-figure \ref{f4} (b) exhibit a similar crowd evacuation process. However, except for the RDUE model, these continuous models initiate the evacuation earlier than the particle model. Moreover, the evacuation efficiencies of these models differ as indicated by the varying slopes of the curves in the figure.

\begin{figure}[htbp]
	\centering
	\subfigure[$t=0$]{
		\includegraphics[width=1.58in]{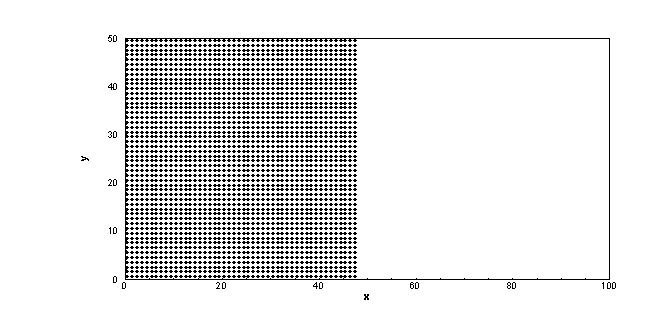}  
	}
	\subfigure[$t=30$]{
		\includegraphics[width=1.58in]{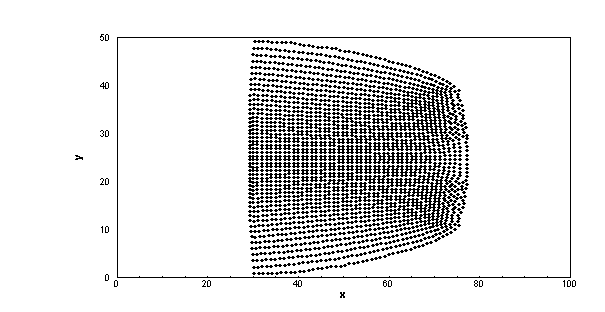} 
	}
	\subfigure[$t=50$]{
		\includegraphics[width=1.58in]{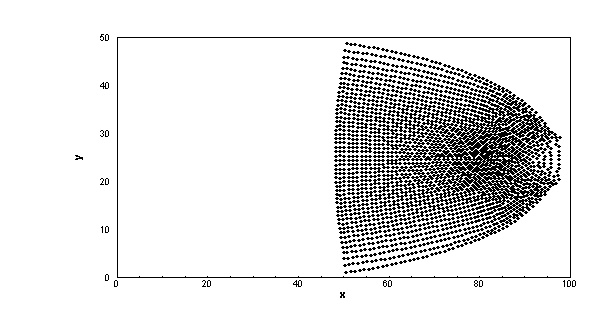} 
	}
    \subfigure[$t=70$]{
	    \includegraphics[width=1.58in]{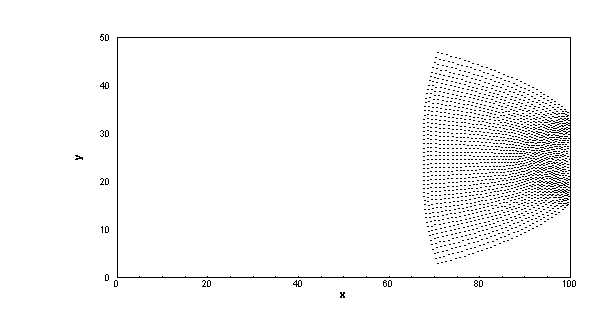}  
    }
	\subfigure[$t=90$]{
		\includegraphics[width=1.58in]{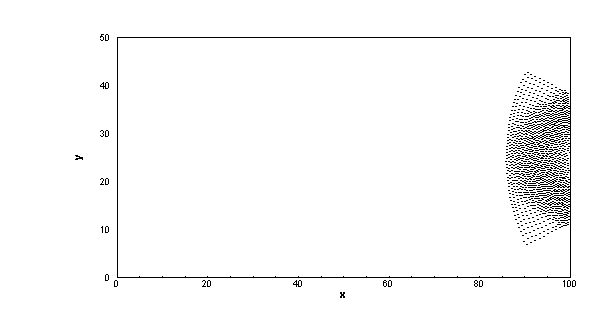} 
	}
	\subfigure[$t=120$]{
		\includegraphics[width=1.58in]{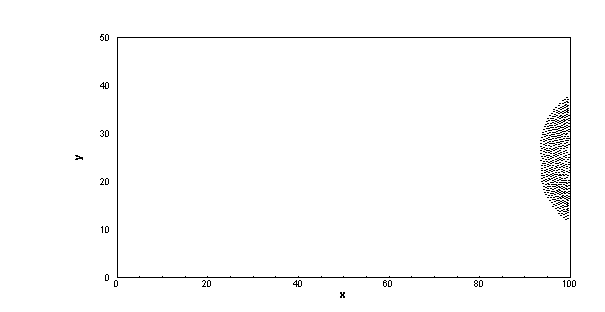} 
	}
	\caption{The evolution of the pedestrian dynamic by the microscopic social force model at different times in Example 2.} \label{f5}
\end{figure}

FIG. \ref{f5} shows the evolution of the pedestrian dynamic by the microscopic social force model at different times in Example 2 (narrow exit). Initially, pedestrians are uniformly distributed without overlapping, and they try to find the optimal path to leave the room based on the RDUE principle. Due to the limitation of the narrow exit, pedestrians gradually gathered at the front of the crowd. As the pedestrians gradually reach the exit, the fan-shaped congestion zone forms at the inside of the exit. Then, the pedestrians queue up to leave the room.

\subsection{Simulations without/with obstacles}
\begin{figure}[htbp]
	\centering
	\subfigure[$t=30$]{
		\includegraphics[width=2.4in]{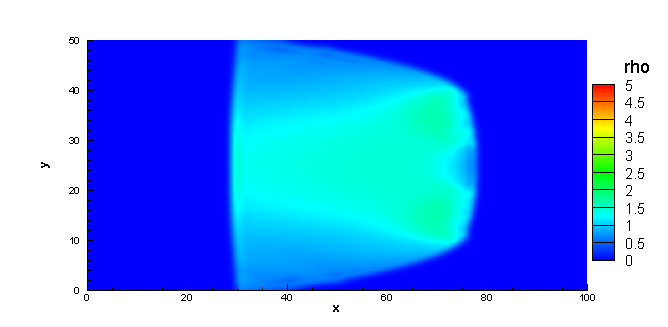}  
		\includegraphics[width=2.4in]{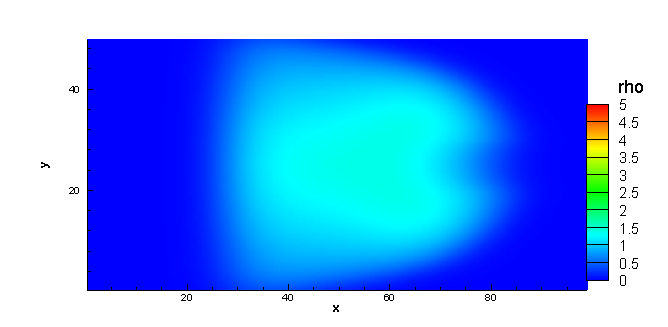}  
	}
	\subfigure[$t=50$]{
	\includegraphics[width=2.4in]{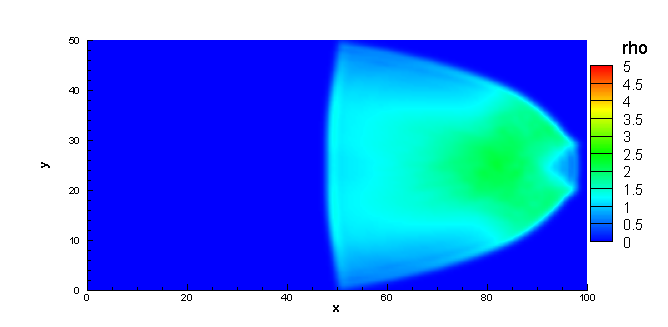}  
	\includegraphics[width=2.4in]{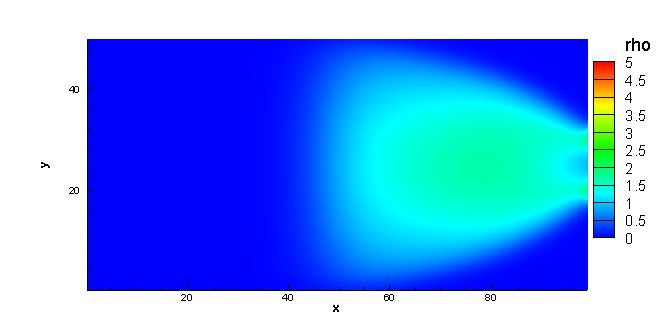}  
    }
	\subfigure[$t=70$]{
		\includegraphics[width=2.4in]{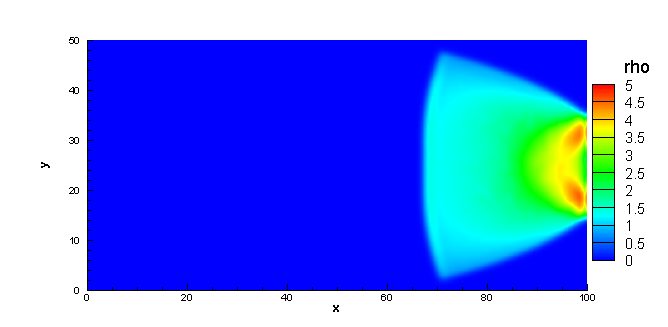}  
		\includegraphics[width=2.4in]{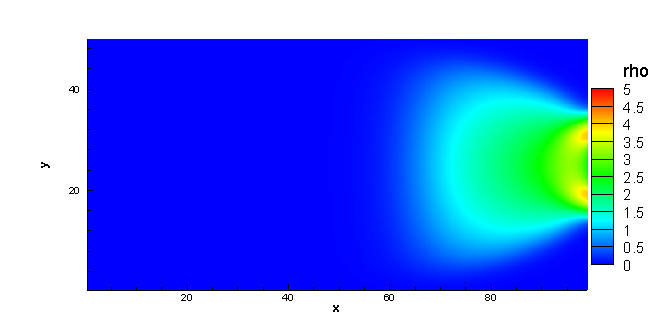}  
	}
	\subfigure[$t=90$]{
		\includegraphics[width=2.4in]{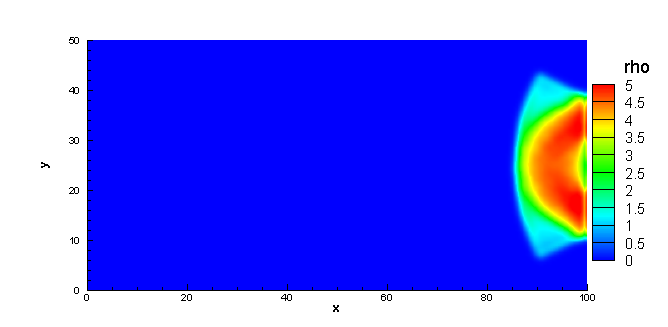}  
		\includegraphics[width=2.4in]{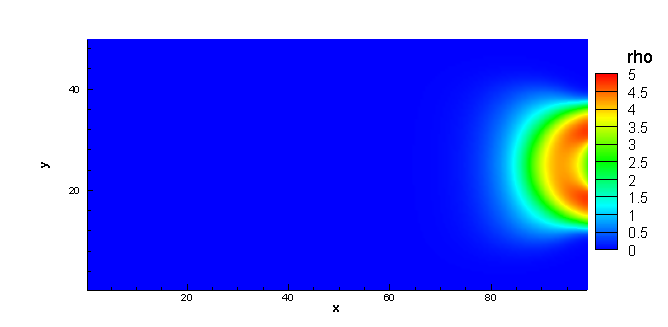}  
	}
	\caption{The temporal and spatial distributions of the density of the social force model (left) and the related macroscopic model (right) in Example 2.} \label{f6}
\end{figure}

\begin{figure}[htbp]
	\centering
	\subfigure[$t=30$]{
		\includegraphics[width=2.4in]{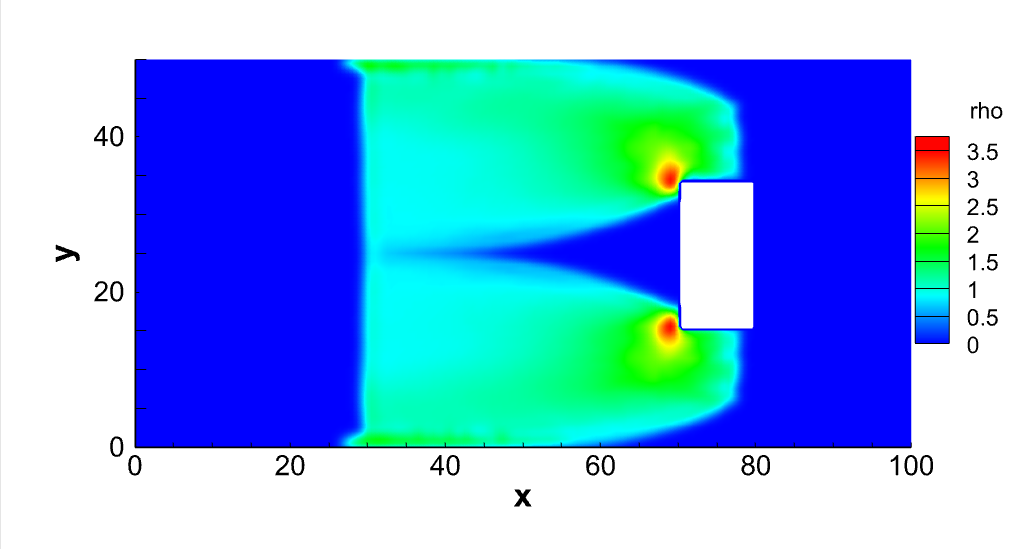}  
		\includegraphics[width=2.4in]{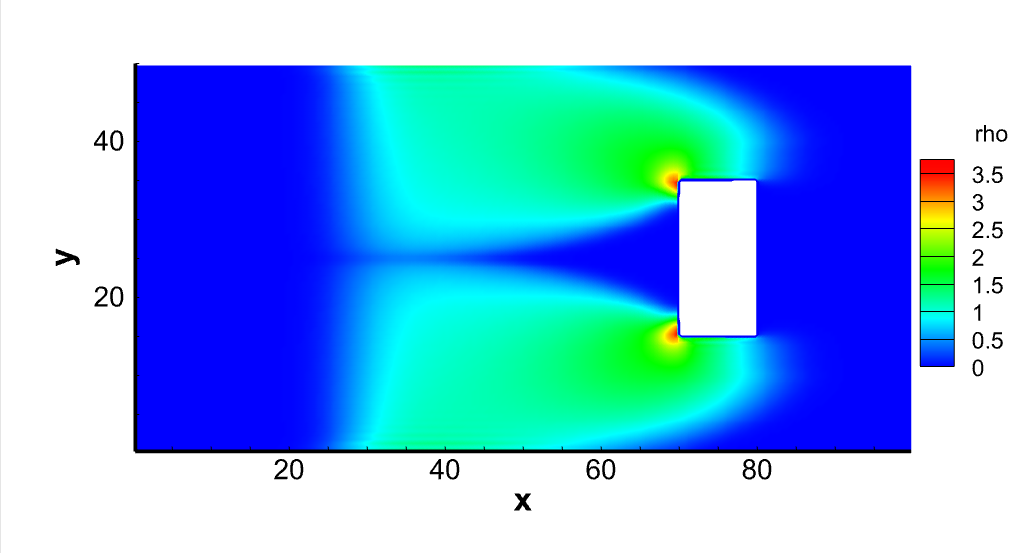}  
	}
	\subfigure[$t=50$]{
		\includegraphics[width=2.4in]{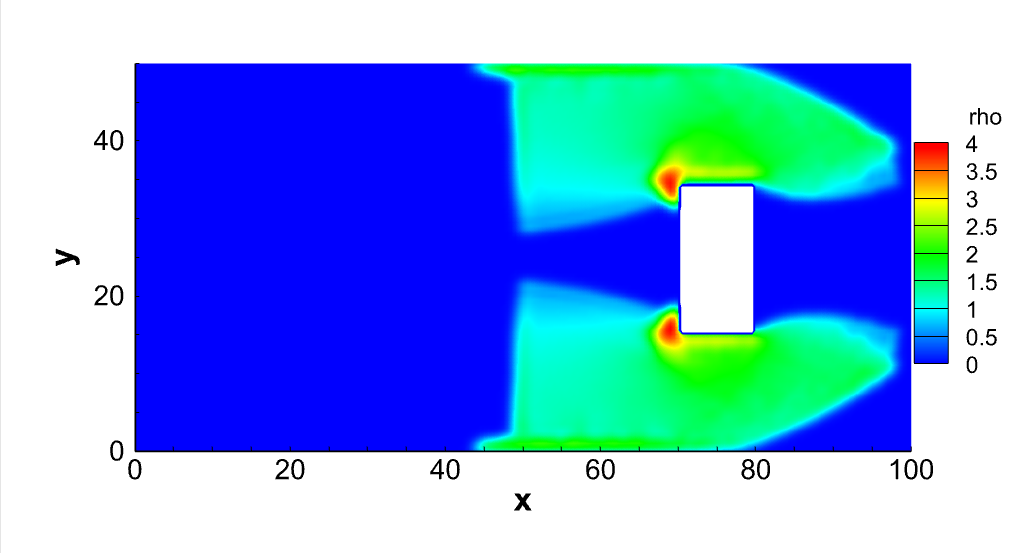}  
		\includegraphics[width=2.4in]{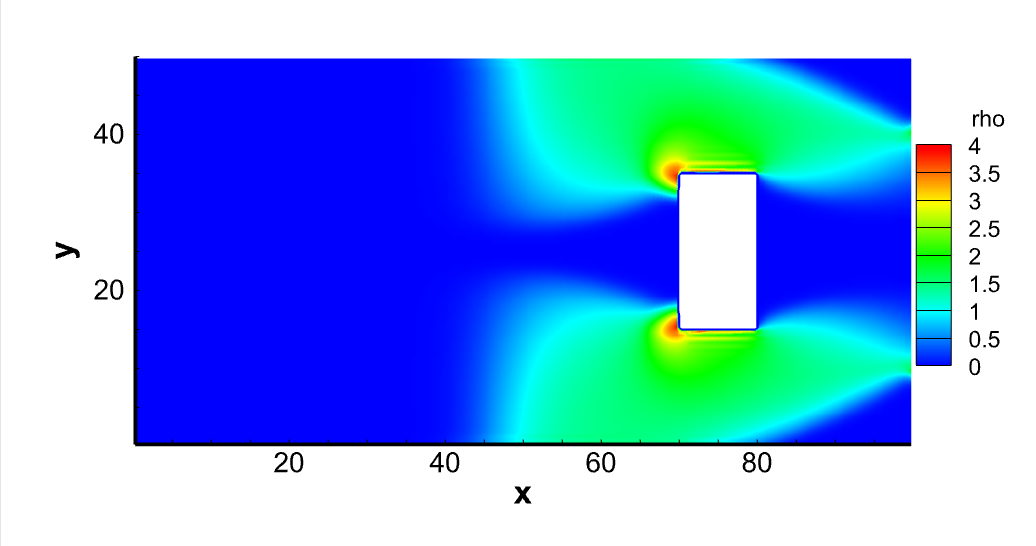}  
	}
	\subfigure[$t=80$]{
		\includegraphics[width=2.4in]{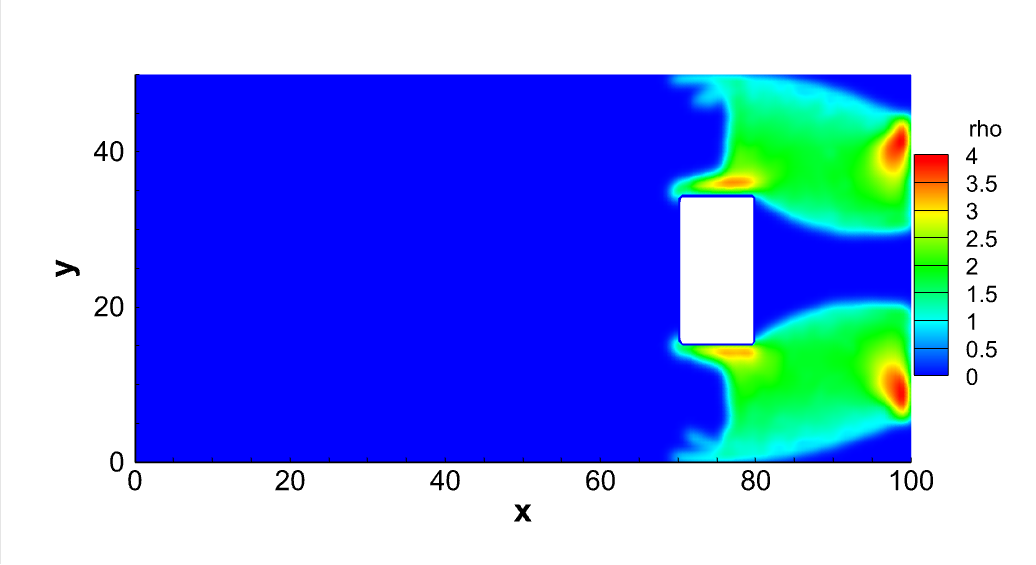}  
		\includegraphics[width=2.4in]{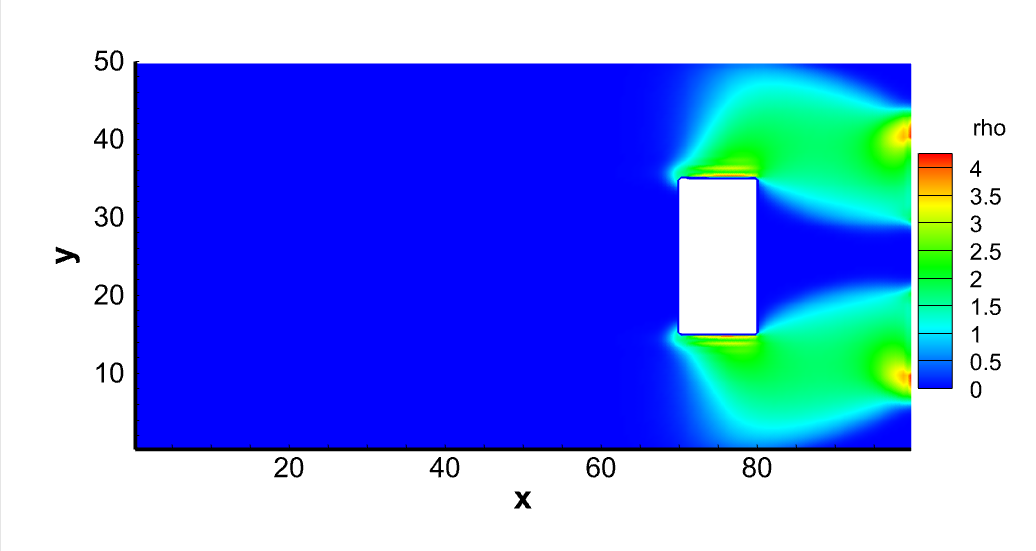}  
	}
	\subfigure[$t=100$]{
		\includegraphics[width=2.4in]{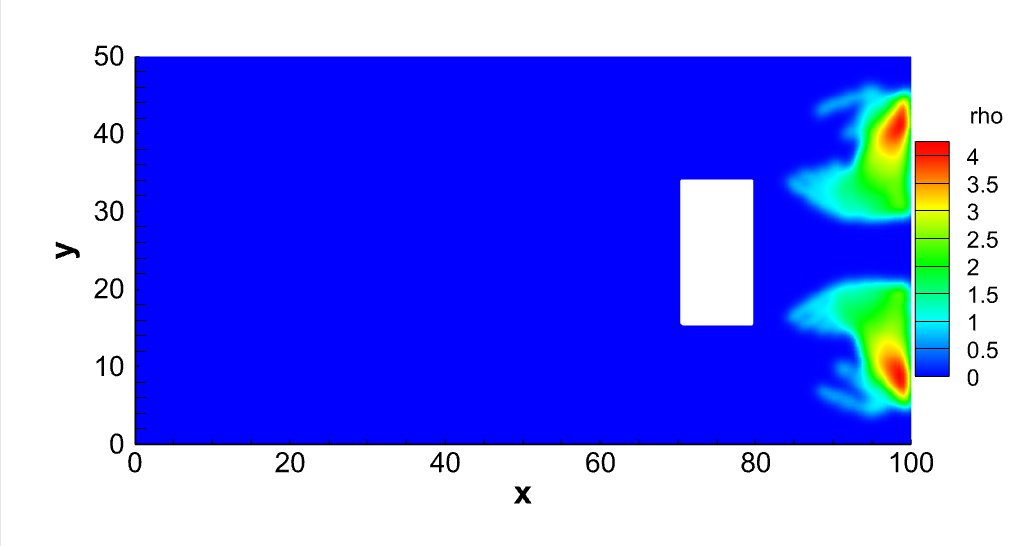}  
		\includegraphics[width=2.4in]{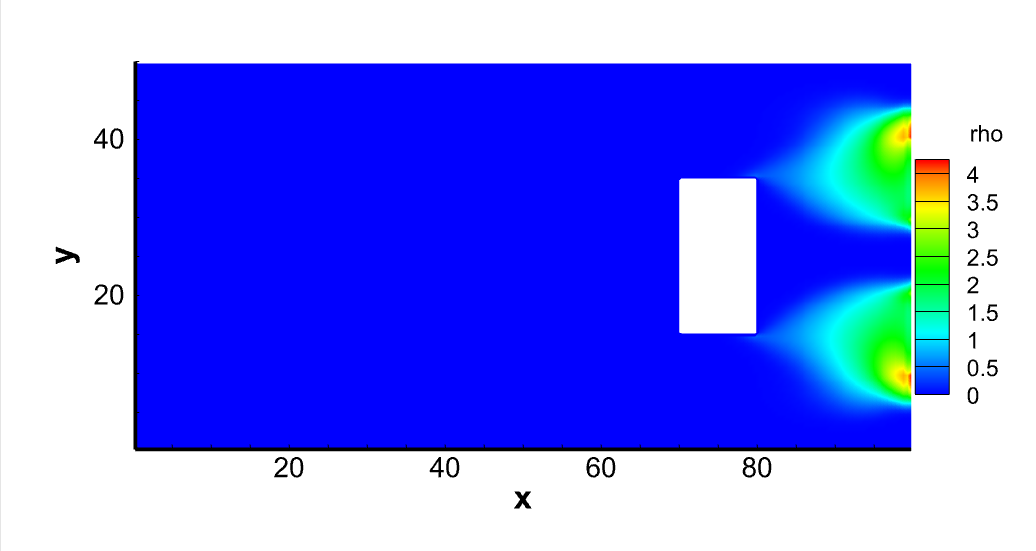}  
	}
	\caption{The temporal and spatial distributions of the density of the social force model (left) and the related macroscopic model (right)  in Example 3.} \label{f7}
\end{figure}

Furthermore, simulations without and with obstacles are used to compare the microscopic and macroscopic model. Figure \ref{f6} and \ref{f7} compare the temporal and spatial evolution of the density for the social force model  (\ref{sf1}) and the corresponding macroscopic model  (\ref{mac1}) in Example 2 and Example 3, respectively. In the particle model, the pedestrian density is computed using equations (\ref{eq3}) and (\ref{eq4}), based on the spatial distribution of individuals. 
By comparing figures at the same time instances, it can be observed that the overall density patterns produced by these two models are highly similar, demonstrating consistent dynamic trends over time and space. These results indicate a close connection between the two models at different levels of description. However, some differences can also be observed, such as variations in the maximum density values and differences in the spatial distribution patterns.
One contributing factor is the potential inaccuracy in the local density estimation formula used in the social force model. More importantly, the derivation of the macroscopic continuum model from the microscopic social force model involves several simplifying assumptions. 

\subsection{Characteristics of pedestrian flow}
\begin{figure}[htbp]
	\centering
	\subfigure[Example 2]{
		\includegraphics[width=2.3in]{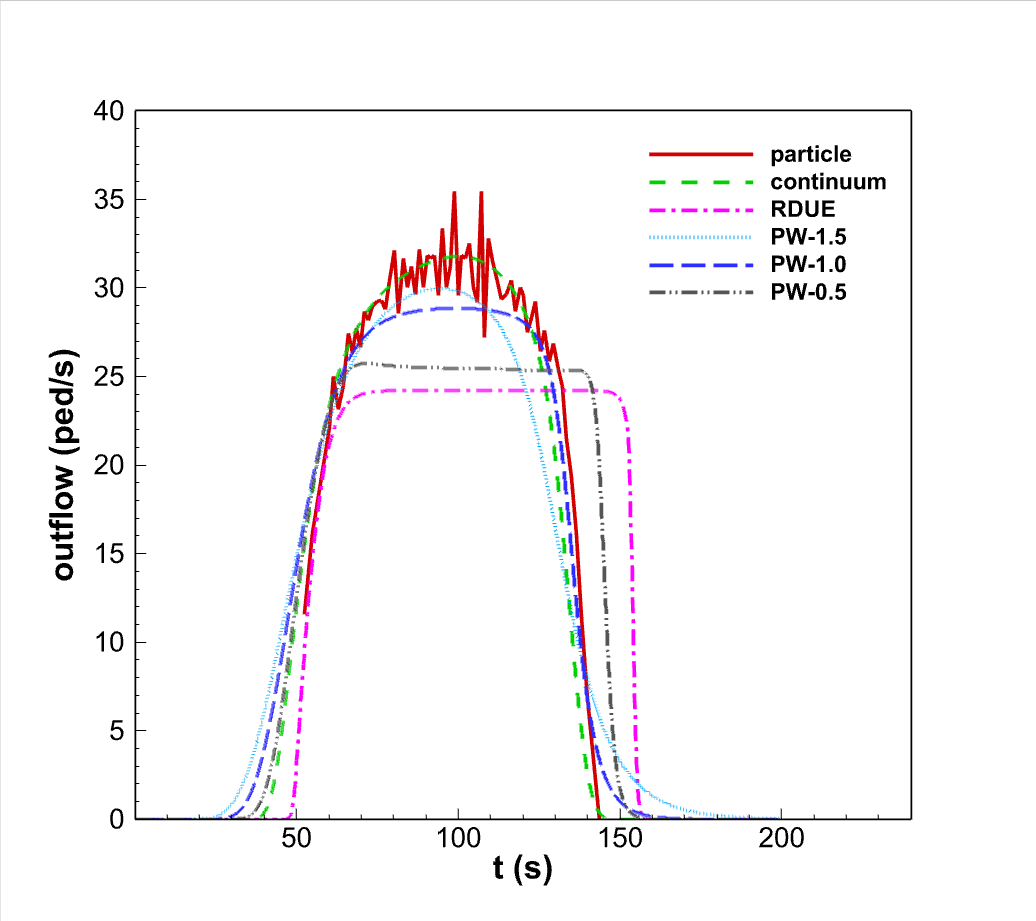}  
	}
	\subfigure[Example 3]{
		\includegraphics[width=2.3in]{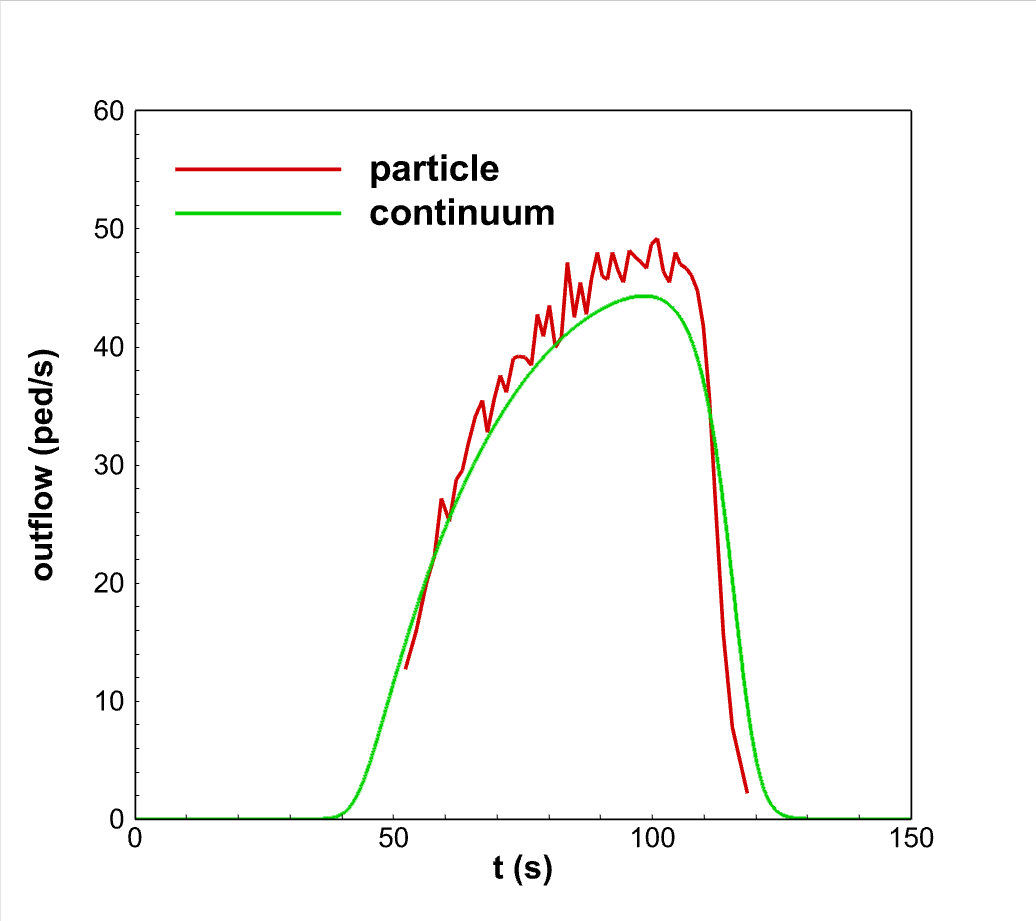}  
	}
	\caption{The outflow through the exit against the simulation time $t$.} \label{f8}
\end{figure}

\begin{figure}[htbp]
	\centering
	\subfigure[Example 1]{
		\includegraphics[width=2.6in]{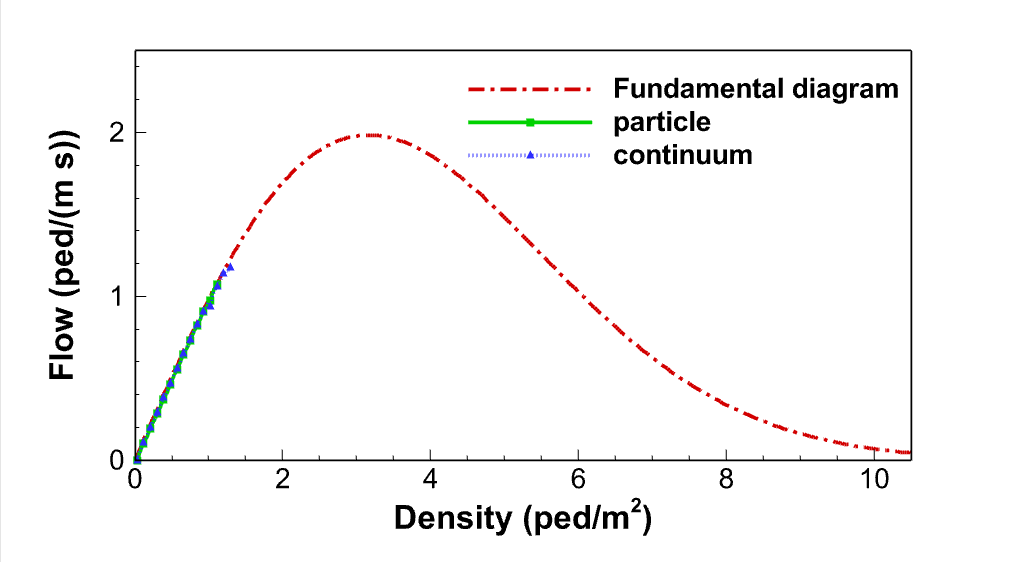}  
	}
	\subfigure[Example 2]{
		\includegraphics[width=2.6in]{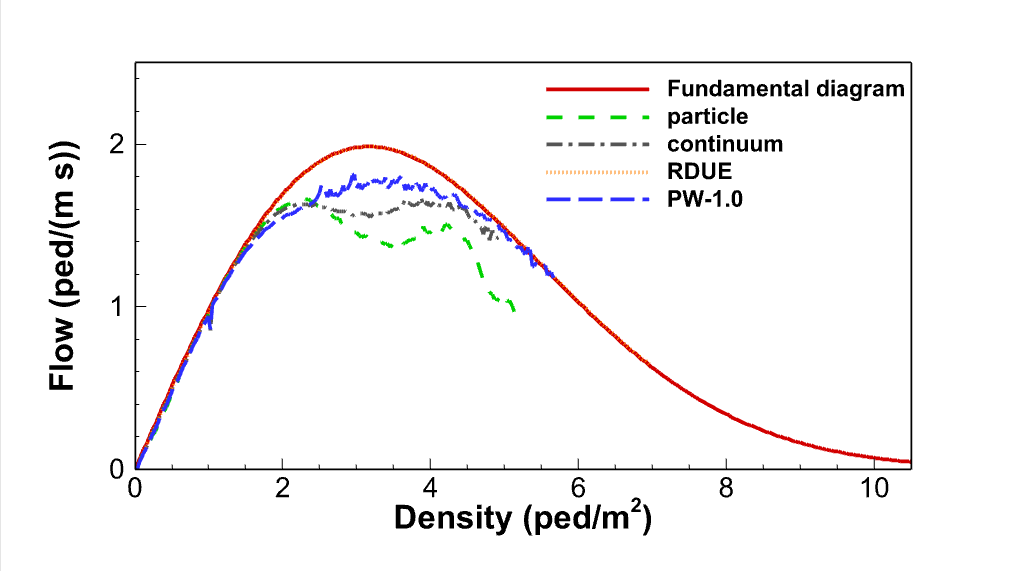}  
	}
	\subfigure[Example 3]{
		\includegraphics[width=2.6in]{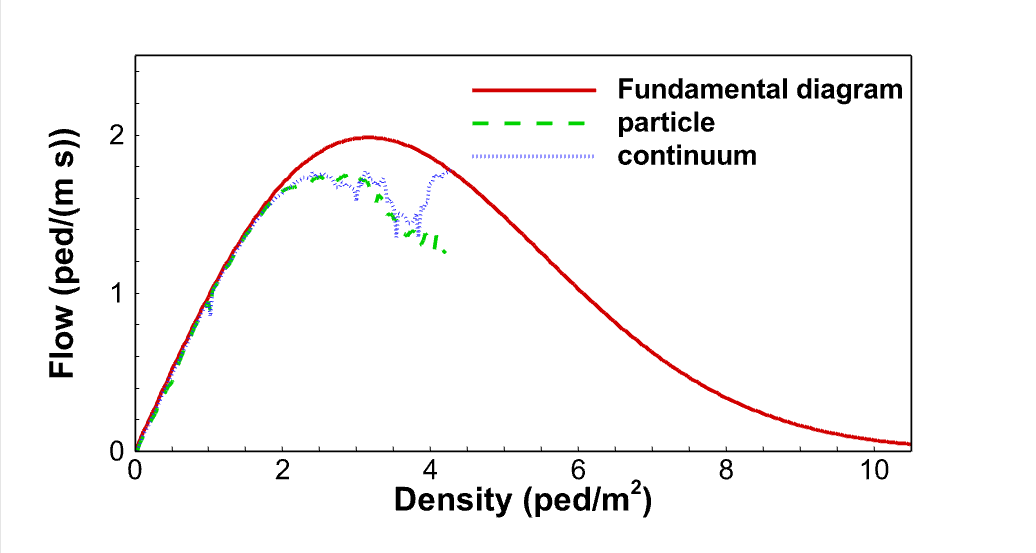}  
	}
	\caption{The local density–flow relationship in particle and continuum models.} \label{f9}
\end{figure}

To study characteristics of pedestrian flow,  the evacuation efficiency of these models are considered, we investigate the time evolution of the outflow at the exit for each model, defined as
\begin{equation}
    f_c(t)=\iint_{\Gamma_c}\rho\bm{u}\cdot\bm{n}ds,
\end{equation}
where $\bm{n}$ is the unit normal vector pointing toward the exit, $\rho$ and $\bm{u}$ is the density and velocity, respectively. In the particle model, the density (local) is computed by equation(\ref{eq3}), and velocity (local) is computed by
\begin{equation}
\begin{aligned}
 \bm{u}(\bm{x},t)=\frac{\sum_{1\leq j\leq N}\bm{v}_{j}w\Big(\|\bm{x}_{j}(t)-\bm{x}\|\Big)}{\sum_{1\leq j\leq N}w\Big(\|\bm{x}_{j}(t)-\bm{x}\|\Big)}.
\end{aligned}
\end{equation}
FIG. \ref{f8} presents the outflow through the exit against the simulation time $t$ under different examples and different models. In this figure, the value of the outflow corresponds to the slope of the curves shown in FIG. \ref{f4}. From these sub-figures, compared to the RDUE model and PW model, the curve of the particle model and the curve of the related continuum model almost completely overlap, thus, our new derived continuum model has a better approximation to the outflow of the particle model.

Finally, the average local flow–density relationship for the evacuation process in the simulation area for the whole evacuation period is shown in FIG \ref{f9}. The data points of local density and flow are obtained by the following procedure:
\begin{itemize}
    \item 
The density and flow magnitudes at each cell (i.e., $(\rho_{i}, f_i))$ are recorded every second during the simulation.
\item
We set a series of densities ranging from 0 to 10.5 ped/m$^2$ with an incremental interval of 0.03 ped/m$^2$. The data points
obtained from the simulation are grouped into these intervals, and the flow is averaged over the data points within each density interval.
\end{itemize}
In FIG. \ref{f9}, the function fundamental diagram is defined as $f_d=\rho U^{\rm e}$, here $U^{e}$ is defined in equation (\ref{eq_desired_speed}). From this figure, our new derived continuum model also has a better approximation to the particle model, compared to other existing continuum models.

Combining the results of density, outflow and the fundamental diagram,  the new derived macroscopic continuum model gives a good approximation of the microscopic social force model. 
Moreover, the macroscopic continuum model is derived from the microscopic social force model. 
So the pressure term and source term in the continuum equation have detailed physical meaning: these terms are derived from the interaction force and desired force. 
We can calibrate the parameters in our derived macroscopic continuum model, such as we can choose the suitable parameters according to the microscopic model for different scenario of pedestrian dynamic problem. 
On the other hand, once we know the connection between the microscopic social force model and the macroscopic continuum model, this can help to use the micro-macro approach to study crowd dynamics.

\section{Conclusion}\label{secConclusion}
This paper derived a continuum model based on the traditional social force model with an optimal route choice strategy, which established a connection between the microscopic model and the macroscopic model.
Thus, the microscopic model can serve as a ground truth for calibrating and interpreting macroscopic parameters, while the macroscopic model enables theoretical analysis and efficient simulations to evaluate and regulate the global forces influencing microscopic dynamics.
At the microscopic level, the traditional social force model was used to model the pedestrian dynamics, which takes into account the interactions of pedestrians, and the desired walking direction is governed by an eikonal equation. Based on the social force model, we considered the time-asymptotic flocking behavior of the social force model. 
Then, we derived a kinetic model at the mesoscopic level and a continuum model at the macroscopic level, in which two different cases are derived: dimensionalized case and hydrodynamic scaling case. 
In the dimensionalized case, we send the number of pedestrians to infinity, and directly derived a mean-field model and the related macroscopic continuum model. 
In the hydrodynamic scaling case, we aim to study the dynamics of the system at large time and space scales, in which according to different assumptions of the range and intensity of the interaction force, we derived four different continuum models.  
Finally, we gave a series of numerical examples to demonstrate the consistency between the social force model and the new related macroscopic model, and show that the macroscopic models can be a powerful tool to study the challenging problems for pedestrian flow. 
In the future work, we will combine the characteristics of the microscopic model and continuum models, perform the model calibration with experimental data, and investigate complex scenarios analytically and numerically.

\section*{Acknowledgement}
The work of L. Yang is supported under Grants NSFC 12271499. The work of H. Yu is supported under Grants NSFC 12271288 and 12571469, and in part by the 111 Project (No. D23017), and Program for Science and Technology Innovative Research Team in Higher Educational Institutions of Hunan Province of China. The work of J. Du is supported by the National Natural Science Foundation of China (Grant No. 12571425), Natural Science Foundation of Shanghai, China (Grant No. 24ZR1417600), and Shanghai Young Academic Program of Eastern Talent Plan.

\bibliographystyle{siamplain}
\bibliography{ref}
\end{document}